\documentclass[11pt]{article}
\usepackage{latexsym}
\usepackage{euscript}

\usepackage{amsmath,amssymb}
\usepackage{xcolor}
\usepackage{textcomp}
\usepackage{color}
\usepackage{graphicx}
\usepackage{amsmath,amssymb,color,graphics,cite,footnote}

\usepackage{epsfig}
\usepackage{epstopdf}

\def\be{\begin{equation}}
\def\ee{\end{equation}}
\def\bea{\begin{eqnarray}}
\def\eea{\end{eqnarray}}

\long\def\symbolfootnote[#1]#2{\begingroup%
\def\thefootnote{\fnsymbol{footnote}}\footnote[#1]{#2}\endgroup}

 \large\normalsize

\usepackage{silence}
\usepackage{soul}
\begin{document}
\thispagestyle{empty}
\vspace*{0.2cm}
\begin{center}
{\Large \bf New Black Hole Solutions of Second and First Order Formulations of Nonlinear Electrodynamics}
\\
\vspace{1.0cm}
{\large Yosef Verbin\footnote{verbin@openu.ac.il}$^{\ast}$, Beyhan Puli{\c c}e \footnote{beyhan.pulice@sabanciuniv.edu}$^{\ast\dagger}$}, Ali \"Ovg\"un \footnote{ali.ovgun@emu.edu.tr}$^{\ddagger}$  and  Hyat Huang\footnote{hyat@mail.bnu.edu.cn}$^{\sharp}$
\\
\vspace{0.5cm}
\small{$^{\ast}$ Astrophysics Research Center, the Open University of Israel, Raanana 4353701, Israel\\
$^{\dagger}$ Faculty of Engineering and Natural Sciences, Sabanc{\i} University, 34956 \.{I}stanbul, Turkiye\\
$^{\ddagger}${Physics Department, Eastern Mediterranean University, Famagusta, 99628 
 via Mersin 10, Turkiye}\\
$^{\sharp}$ College of Physics and Communication Electronics, Jiangxi Normal University, Nanchang 330022, China}
\end{center}


\begin{abstract}
Inspired by the so-called  Palatini formulation of General Relativity and of its modifications and extensions, we consider an analogous formulation of the dynamics of a self-interacting gauge field which is determined by non-linear extension of Maxwell's theory, usually known as nonlinear electrodynamics. In this \textit{first order} formalism the field strength and the gauge potential are treated, \textit{a priori} as independent, and, as such, varied independently in order to produce the field equations. Accordingly we consider within this formalism alternative and generalized non-linear Lagrangian densities, some of them of a new kind which gives up the restriction of equivalence to second order Lagrangians.  Several new spherically-symmetric objects are constructed analytically and their main properties are studied. 
 The solutions are obtained in flat spacetime ignoring gravity and for the self-gravitating case with emphasis on black holes. As a background for comparison between the first and second order formalisms, some of the solutions are obtained by the conventional second order formalism, while for others a first order formalism is applied. Among the self-gravitating solutions we find new families of black holes and study their main characteristics. Some of the flat space solutions can regularize the total energy of a point charge and a subset of them exhibit also finite field strength and energy density, although their black hole counterparts are not regular.
\end{abstract}

\newpage
\tableofcontents


\newpage
\section{Introduction}
\setcounter{equation}{0} 
\setcounter{footnote}{0}
\subsection{Background}

The so-called \textit{Palatini formulation} of General Relativity\footnote{The variational principle according to which the metric and the affine connection are varied independently, is commonly known as the “Palatini formulation”. it is argued that this term is not fully justified from a historical point of view, but we will stick to common practice and bypass this historical issue.} is known for a century as an equivalent first order formulation of General Relativity (GR)  and as a non-equivalent route for first order formulations of  non-linear extensions of GR such as $f(R)$  or scalar-tensor theories. Analogously, we wish to consider a variational formulation of electrodynamics and its non-linear extensions where the field strength and the gauge potential are treated, \textit{a priori} as independent, and, as such, varied independently. Accordingly we will present and consider alternative  action functionals, with alternative first order Lagrangian densities and study the main consequences.

We will see that the analogous (first order) construction for the electromagnetic field has similar characteristics as for gravity: it is equivalent to the second order ordinary Maxwell formalism in the linear theory, becomes inequivalent when non-linearities appear, but unlike the GR case, there is an intermediate domain in the ``space of theories'' where both formulations are equivalent i.e., their field equations are equivalent. This intermediate domain is already well-known thanks to the influential work of Plebanski \cite{Plebanski1970} and has been used in studying theories of non-linear electrodynamics (NLED) and their self-gravitating versions with applications like BH solutions and especially regular BHs. 

However, all studies which are known to us were confined to the intermediate domain where the equivalence between first and second order formulations was exploited in NLED theories usually defined through second order, to study aspects that are easier to treat in first order. No existing purely first order construction of an electromagnetic theory which is independent of a second order ``shadow'' seems to exist in the literature. 

In this paper, we present such a pure first order formulation of NLED without insisting on equivalence with second order. We do it after gaining a perspective from a survey of the equivalent linear theories in first and second order formulation, and the much less trivial correspondence between theories that belong to the intermediate domain of the NLED theories.

\subsection{Second order formulation of Maxwell electrodynamics}
\label{SecondOrderMax}

The standard covariant formulation of electrodynamics relies on the Maxwell Lagrangian density $\mathcal{L}_{\rm M}=-\frac{1}{4}F^{\mu\nu} F_{\mu\nu}-J^\mu A_\mu$ when $J^\mu$ is the 4-current.

In this formulation, it is \textit{assumed} from the outset that the fundamental dynamical variable is the 4-potential $A_\mu$, which relates to  the  Maxwell-Faraday 2-form as $F=dA$ or in components
\begin{equation}
F_{\mu\nu}=\partial_\mu A_\nu-\partial_\nu A_\mu \ .
\label{bianchi0}
\end{equation}
Then, varying the action with respect to the fundamental variable $A_\mu$ gives
\begin{equation}
\nabla_\mu F^{\mu\nu}=J^\nu \ ,
\label{nonhom}
\end{equation}
where $\nabla_\mu$ is the spacetime covariant derivative associated to the Levi-Civita connection of the metric\footnote{we use the $(+---)$ signature and units where the speed of light = 1. $g = det(g_{\mu\nu})$.}  .

Eqs.~\eqref{bianchi0} or its equivalent $F=dA$  imply the standard source-free (homogeneous) Maxwell equations $dF=0$  (\textit{aka} Bianchi identities), or in components  $\partial_{[\lambda}F_{\mu\nu]}=0$. An alternative form in terms of the dual field-strength $^{*}F^{\mu\nu} = \epsilon^{\mu\nu\rho\sigma}F_{\rho\sigma}/2\sqrt{-g}$  is also in use:
\begin{equation}
\nabla_{\mu} \,^{*}F^{\mu\nu}=0 \ .
\label{bianchi}
\end{equation}
This is a second order formalism, as the equations of motion are naturally second order differential equations in terms of the fundamental vector potential $A_\mu$.
\subsection{First order formulation of Maxwell electrodynamics}
\label{FirstOrderMax}

As mentioned above, inspired by the Palatini formulation of GR we wish to consider a formulation of electrodynamics where $F_{\mu\nu}$ and $A_\mu$ are treated, \textit{a priori} as independent, and, as such, varied independently. Accordingly we consider the alternative action, with an alternative Maxwell Lagrangian density
\begin{equation}
\mathcal{L}_{\rm M}^{(1)}=\frac{1}{4}F_{\mu\nu}F^{\mu\nu}-\frac{1}{2}F^{\mu\nu}(\partial_\mu A_\nu-\partial_\nu A_\mu)-J^\mu A_\mu \ .
\label{MaxwellFirst}
\end{equation}
The superscript $(1)$ refers to first order formalism, where nothing is assumed about the relation between $A_\mu$ and $F_{\mu\nu}$, so these two tensors are varied independently in the corresponding action:
\begin{description}
\item[$\bullet$] Varying with respect to $A_\mu$ yields the source-full Maxwell equations~\eqref{nonhom};
\item[$\bullet$] Varying with respect to $F_{\mu\nu}$ yields the equality $F_{\mu\nu}=\partial_\mu A_\nu-\partial_\nu A_\mu$ which is equivalent to the source-free Maxwell equations~\eqref{bianchi}.
\end{description}
Thus, Eq.\eqref{MaxwellFirst} establishes a first order formulation of a field theory which is equivalent to the Maxwell theory\cite{Lanczos1949-1970, Schwinger1953}. See also \cite{Faddeev1969} for an analogous result for Yang-Mills theories. This is reminiscent of what occurs in GR, for which the standard second order formulation and the Palatini formulation are equivalent. It is well known that this ceases to be the case in higher order curvature theories - see $e.g.$~\cite{Olmo:2011uz}. Thus we are motivated to consider further interaction terms to assess the distinction between  the two formulations in the gauge-theory case. In Sec. \ref{PNLED} we will see that the analogous situation for NLED is indeed realized.

\subsection{Plan of this paper} \label{PaperPlan}  

The main aim of this paper is to present a new family of NLED theories  in first order formalism which is free of the generally imposed condition of equivalence to NLED in second order. Then we will choose a specific model and study some of its basic features and characteristics, like the fields of electric point charges, their energy density and total energy (which turns out to be finite) and then moving to consider self-gravitating solutions, BHs and their properties. 

This will be done in Sec.\ref{PNLED}. But before that we give some background in order to get some perspective and appreciate the context of these results. We will therefore start in Sec.\ref{SecondOrderNLED} with a general introduction to NLED, which will be followed by Sec.\ref{GHE} with a study if a simple NLED model in second order formulation in flat spacetime and then coupled to gravity. In  Sec.\ref{FirstOrderNLEDversion1} we will summarize the way to construct a first order NLED Lagrangian from a given second order one such that both formulations will correspond to the same theory. In order to demonstrate the reverse direction, we will study in  Sec.\ref{FirstOrder-Poly}  a system in first order formalism which is similar to the one studied in second order in Sec.\ref{GHE}, and examine the relations between them, and the general conditions for equivalence which is not a-priory guaranteed. 

Finally, in Sec.\ref{PNLED}, we suggest a direction to go further in this way and construct new theories which are formulated from the outset in first order giving up any considerations of equivalence  with second order NLED. An application of this formalism in a different context of cosmological inflation with an additional scalar field was reported recently in Ref.~\cite{Rasanen-Verbin2022}.

\section{Nonlinear electrodynamics in  second order formalism} \label{SecondOrderNLED}
\setcounter{equation}{0}


Nonlinear extensions of Maxwell theory have a long history \cite{Sorokin2022} of at least 90 years since the works of Born and Infeld \cite{BornInfeld}   and Euler, Kockel and Heisenberg\cite{Euler-Kockel,Heisenberg-Euler} in the 1930s. Both of these theories may be viewed as special cases of the general family of Lagrangians of Nonlinear Electrodynamics  depending on the two Maxwell invariants such that
\begin{equation}
\mathcal{L}^{(2)}_{\rm NLED}= \frac{1}{4}f(X,\Xi) -J^\mu A_\mu  \;\;,\;\;\; X= F_{\mu\nu}F^{\mu\nu} \;\;\;,
\;\;\Xi= F_{\mu\nu}\,^{*}F^{\mu\nu}
\label{LagNLED2-general}.
\end{equation}
The current density $J_\mu$ is kept just to keep track of the modifications of the field  equations. In most concrete calculations in this paper it will be omitted.
The field equations of this theory are as usual, split to the source free equations \eqref{bianchi} which are unaffected by the nonlinear extension, and the source-full equations which generalize to
\begin{equation}
-\nabla_{\mu} \left( f_{X}F^{\mu\nu}   +  f_{\Xi}\,^{*}F^{\mu\nu} \right) =  J^\nu
\label{NLED-FEqs}
\end{equation}
where $f_{X}$ and $f_{\Xi}$   are the partial derivatives of $f(X,\Xi)$. The Maxwell limit is obtained  if $f_{X} \rightarrow -1$ and $f_{\Xi} \rightarrow 0$ for weak fields, together of course with $f(X,\Xi) \rightarrow 0$.

Since we will be interested in self-gravitating objects in this kind of theories we complete this short general presentation by writing down the energy-momentum tensor of the self-interacting gauge field (discarding the current term) which is obtained, as usual, by varying the action with respect to $g_{\mu\nu}$:
\be
T_{\mu\nu}= f_{_X}F_{\mu}^{\ \alpha}F_{\nu\alpha} +\frac{1}{4}\left(\Xi f_{_\Xi} -f(X,\Xi)\right)g_{\mu\nu}\ .
\label{Tmunu2nd}
\ee

Specifically, the Born-Infeld (BI) and Heisenberg-Euler (HE)  Lagrangians are given by
\begin{equation}
f_{BI}(X,\Xi)=4b^2 \left(1-\sqrt{1+\frac{1}{2b^2} X -\frac{1}{16b^4} \Xi^2}\right) \;\;,\;\;\; f_{HE}(X,\Xi)= -X+a_1 X^2 +a_2 \Xi^2
\label{BI+HE},
\end{equation}
where $b$, $a_1$ and $a_2$ are positive parameters. Actually, $a_1$ and $a_2$ have definite values calculated in \cite{Euler-Kockel,Heisenberg-Euler} and also by other methods by Schwinger \cite{Schwinger1951} from the QED Lagrangian. We have to clarify further that the function $f_{HE}(X,\Xi)$ does not represent the full nonlinear Lagrangian of Ref.\cite{Heisenberg-Euler}, but rather the simpler version obtained by Euler and Kockel \cite{Euler-Kockel}. For the sake of historical accuracy, we should use ``HEK'' or ``EK'' to include Kockel, but we will follow most authors, write ``HE'', and refer to this Lagrangian as ``Heisenberg-Euler''.

These nonlinear Lagrangians originated from a variety of directions and motivations 
in addition to the obvious one already alluded to, of obtaining quantum-mechanical corrections to the interaction of photons with matter \cite{Sorokin2022,Euler-Kockel,Heisenberg-Euler,Schwinger1951,Bialynicka-Birula1970,Adler1971,Drummond1979}. These are:  mitigating the singular fields and diverging field energies of point charges in the Maxwell theory \cite{Sorokin2022,BornInfeld, Plebanski1970, Kruglov2014,CostaEtAl2015}, or studying the effects of vector field nonlinearities on the well-known Maxwell-Einstein  black holes \cite{Pellicer+Torrence1969,deOliveira1994,Yajima+Tamaki2000,RuffiniWuXue2013,Kruglov2014,Kruglov2015BH,Kruglov2017,Breton+Lopez2021} and using them in order to construct regular black holes \cite{Ayon-Beato1998,Bronnikov2001,Dymnikova2004,BalartVagenas2014,Bambi:2013ufa,Fan:2016hvf,Li:2024rbw,Ovgun:2019wej,Nicolini:2005vd}. 

Bardeen \cite{Bardeen1968} was the first to propose a regular black hole (BH) metric with horizons and no singularity by introducing a central matter core and replacing the Schwarzschild mass with a radius-dependent function, thereby eliminating singularities \cite{Bardeen1968}. The resulting spherically symmetric metric, known as the Bardeen regular black hole, violates the strong energy condition. Additionally, other similar spherically symmetric regular black hole models have also been proposed \cite{Dymnikova:1992ux,Dymnikova:2003vt,Hayward:2005gi}. Later, Ayon-Beato and Garcia \cite{Ayon-Beato1998,Ayon-Beato:1999kuh} demonstrated that such solutions could also be constructed within GR in the presence of a nonlinear electrodynamic field and identified a magnetic monopole of a specific NLED model \cite{Ayon-Beato2000} as a possible source of the Bardeen BH. The source of the Hayward BH was similarly identified \cite{Fan+Wang2016} as a magnetic monopole of another NLED model.

NLED Lagrangians of the BI family were also found in the framework of string theory as effective low energy descriptions of open strings \cite{Fradkin+Tseytlin1985,Tseytlin2000} and studies of the point charge fields were done also  from this point of view \cite{Tseytlin1995,Pasarin+Tseytlin2020}.  NLED cosmological models were also suggested \cite{Kruglov2015Cosm,Garcia-Salcedo+Breton2000,ElizaldeEtAl2003,NovelloEtAl2003,Ovgun:2016oit,Benaoum:2022uta,Otalora:2018bso,Ovgun:2017iwg}.


\section{Generalized Heisenberg-Euler Theory}
\label{GHE}
\setcounter{equation}{0}

We begin with a simple generalization of the Heisenberg-Euler (GHE) Lagrangian to higher (integer) powers of $X$ and $\Xi$ which we write in a slightly more convenient parametrization:
\begin{equation}
 f_{GHE}(X,\Xi)= -X+\frac{2\gamma}{n} X^n +\frac{2\beta}{n} \Xi^n
\label{fGHE_n},
\end{equation}
where $\gamma$ and $\beta$ are real parameters. We will concentrate on spherically-symmetric electric fields first in flat spacetime and then study the self-gravitating case.

We thus start with writing explicitly the field equations of the GHE Lagrangian. The homogeneous field equations keep their linear Maxwellian form, but the source-full ones are modified:
\begin{equation}
\nabla_\mu\left[\left(1-2\gamma X^{n-1}\right)  F^{\mu\nu}  -2\beta  \Xi^{n-1} \,^{*}F^{\mu\nu} \right]=J^\nu \ .
\label{FEqsNLED-n}
\end{equation}
The energy-momentum tensor of this  GHE  vector field gets the following form (see \eqref{Tmunu2nd}):
\begin{equation}
T_{\mu\nu}=-\left(1- 2\gamma  X^{n-1} \right)F_{\mu\alpha}F_{\nu}^{\ \alpha} +
\frac{1}{4}\left( X - \frac{2\gamma}{n}  X^n + \frac{2(n-1)\beta}{n} \Xi^n\right)g_{\mu\nu}
 \ .
\label{TmunuGEH}
\end{equation}

\subsection{Flat Space GHE Spherically-Symmetric Electrostatic Solutions}
\label{GHESol-Flat}

\begin{figure}[t!]
\begin{center}
{\includegraphics[width=0.49\textwidth]{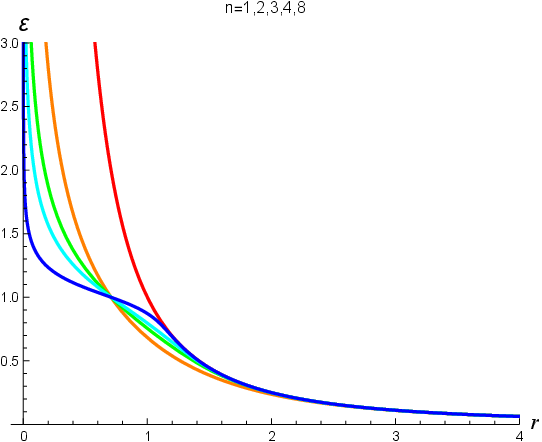}
\includegraphics[width=0.49\textwidth]{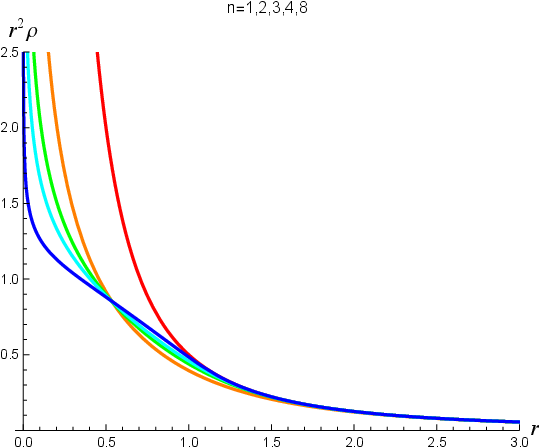}}
\caption{\small{Plots of the dimensionless electric field  $\mathcal{E}(r)$ and the energy density as $r^2 \rho (r)$ of a point charge in 2nd order formalism NLED  for several values of the power $n$:  $n= 1$(linear theory with $\gamma=0$), $2,\,3,\,4,\,8$. A ``spectroscopic'' color code is used such that color frequency increases with $n$. We use this convention throughout the paper in most figures for other relevant parameters.}}
\label{FigProfilesNLED2}
\end{center}
\end{figure}

The first situation to consider is the electrostatic field of a point charge, $Q$. Assuming static and spherical symmetry, we will have a single component of the gauge potential, $A_{t}(r)$ and accordingly a single field component $F_{tr}(r)$. The single equation which remains of Eqs \eqref{FEqsNLED-n} is easily integrated  for general $n$ and $F_{tr}(r)$ is trivially found to satisfy the following equation:
\begin{equation}
F_{tr}+(-2)^n\gamma F_{tr}^{2n-1} =\frac{Q}{r^2} \ .
\label{EqFtrSecondOrderGen_n}
\end{equation}
It is obvious that a well-behaved electric field (for example defined for any $r>0$) requires that the sign of  $\gamma$ will be correlated with the power $n$ such that $(-1)^n\gamma$ will be always positive, so we will impose this condition always for this system. In this case the left-hand-side of Eq. \eqref{EqFtrSecondOrderGen_n} is monotonic in $F_{tr}$ and can be always assumed to be non-negative (together with $Q>0$), so it is easy to infer the $r$-dependence of $F_{tr}$ both asymptotically and near the origin. Near the origin the second term is dominant, so we obtain the behavior  $F_{tr}(r) \simeq 1/r^{2/(2n-1)}$. The asymptotic behavior is seen easily to be Coulombic. Fig.\ref{FigProfilesNLED2}(Left) depicts (directly from Eq.~\eqref{EqFtrSecondOrderGen_n}) the electric field profiles for several values of the parameter $n$.

Next we write the components of the energy-momentum tensor for a point charge, or rather for the more general case of a spherically-symmetric electric field in vacuum:
\bea
T_t^t=T_r^r=  \frac{1}{2}F_{tr}^2+\frac{(2n-1)(-2)^{n} \gamma}{2n} F_{tr}^{2n} \nonumber \\
T_{\theta}^{\theta}=T_{\phi}^{\phi}=-\frac{1}{2}F_{tr}^2-\frac{(-2)^{n} \gamma}{2n} F_{tr}^{2n}
\label{TmunuSphericalElField-NLED2}
\eea
Fig.\ref{FigProfilesNLED2}(Right) depicts the energy density of the field.  The total energy is obtained by integrating the energy density $\rho=T_{t}^t$ with the additional condition $(-1)^n\gamma >0$.
Both $F_{tr}(r)$ and $r^2\rho(r)$ decrease monotonically with $r$, but as mentioned above, the rate of decay depends on $n$, with higher values of $n$ leading to more pronounced nonlinearity near $r=0$. We will see that in this system, although the electric field diverges at the origin, the total field energy converges.

Since the explicit expression for  $F_{tr}(r)$ is unavailable, the integration will be performed on $F_{tr}$. It turns out that the expressions will be more compact and easier to manipulate if the self-interaction parameter $\gamma$ is traded for an electric field parameter $\mathfrak{E}$ defined by 
$(-2)^n \gamma =1/\mathfrak{E}^{2(n-1)}$, such that a dimensionless electric field will be defined as $\mathcal{E}=F_{tr}/\mathfrak{E}$. We get therefore:
\begin{equation}
E=4\pi\int_{0}^{\infty}  r^2\rho dr=-4\pi\mathfrak{E}^2 \int_{0}^{\infty}  r^2 (\mathcal{E}) \left[ \frac{1}{2}\mathcal{E}^2+\frac{(2n-1)}{2n}\mathcal{E}^{2n}  \right] \frac{dr}{d\mathcal{E}}  d\mathcal{E} \ ,
\label{TotalEnergy2ndOrderNLEDSphGen_n}
\end{equation}
where the inverse function $r(\mathcal{E})$  and $ dr/d\mathcal{E}$ are obtained directly from \eqref{EqFtrSecondOrderGen_n}, assuming $Q>0$ without loss of generality. Thanks to the electric field behavior at both integration limits, the energy is always finite (see also \cite{CostaEtAl2015}) for all powers larger or equal 2 in which we are interested in this work. This result demonstrates that a field energy of a NLED point charge may be finite even if the field itself diverges. It may even be regarded as an advantage over the BI theory, and it is remarkable that it was not reported  (to the best of our knowledge) before 2015 \cite{CostaEtAl2015}.

Fortunately, the integration can be performed explicitly, but even without that,  we can identify separately the dependence on $Q$ and
$\mathfrak{E}$  (or $|\gamma|$) by elementary scaling inspection of Eq. \eqref{TotalEnergy2ndOrderNLEDSphGen_n} and the rescaled form of \eqref{EqFtrSecondOrderGen_n}: $\mathcal{E}+\mathcal{E}^{2n-1} = Q/\mathfrak{E} r^2$. 
Thus we can conclude that the function $E(\mathfrak{E},n,Q)$ has the following behavior:
\begin{equation}
E(\mathfrak{E},n,Q)= Q^{3/2} \mathfrak{E}^{1/2}  v(n) \ ,
\label{TotalEnergy2ndOrderNLEBehavior}
\end{equation}
where $v(n)$ is a function of $n$ only. As a check we notice that the total field energy diverges at the limit $\gamma \to 0$ or $\mathfrak{E}\to \infty$ which corresponds to the Maxwell linear theory. Explicit integration (after two integrations  by parts and other manipulations) yields the following result (contained already for $n=2$ in different forms and contexts in \cite{deOliveira1994,Yajima+Tamaki2000}):
\begin{equation}
E(\mathfrak{E},n,Q) = \frac{8\sqrt{\pi } Q^{3/2} \mathfrak{E}^{1/2} }{3}
  \; \cdot \; \frac{\Gamma\left(\frac{1}{4 (n-1)}\right) \Gamma \left(\frac{6 n-7}{4    (n-1)}\right)}{n-\frac{3}{2} }
\label{TotalEnergy2ndOrderNLED-Analytic} .
\end{equation}
It is now easy to see that the field energy $E(\mathfrak{E},n,Q)$ decreases with $n$ approaching the asymptotic value of $16\pi  Q^{3/2} \mathfrak{E}^{1/2} /3$.

\subsection{Self-Gravitating GHE Spherically-Symmetric Electrostatic Solutions}
\label{GHESol-Grav}
It is both interesting and important to look for black hole (BH) solutions of  NLED theories. First we present and study the solutions of the  GEH theory in the 2nd order formalism, i.e. the one which is defined by  Eq.~\eqref{fGHE_n} above. This has some overlap with previous works \cite{deOliveira1994,Yajima+Tamaki2000,RuffiniWuXue2013,Breton+Lopez2021}, but goes beyond them in other respects. In particular, all previous analytic solutions were found for low values of the power $n$, usually, $n=2$. We present here (and in fact throughout the whole paper) analytic solutions for all $n$ and the resulting expressions for physical characteristics like mass and temperature. 

Minimal coupling of gravity to this GEH Lagrangian density is obtained by adding the Einstein-Hilbert Lagrangian (omitting the current term)
\begin{equation}
\mathcal{L}=\frac{1}{2\kappa}R-\frac{1}{4}F_{\mu\nu}F^{\mu\nu}+\frac{\gamma}{2n}(F_{\mu\nu}F^{\mu\nu})^n +\frac{\beta}{2n}(F_{\mu\nu}\;^{*}F^{\mu\nu})^n\; ,
\label{EqSelf-Grav-NLED-2ndOrder}
\end{equation}
which produces Einstein's equations $G_{\mu\nu}=-\kappa T_{\mu\nu}$ with the energy-momentum tensor given in Eq.~\eqref{TmunuGEH}. We use the notation: $R^\kappa_{\;\lambda\mu\nu} = \partial_{\nu} \Gamma_{\mu\lambda}^{\kappa}- \partial_{\mu} \Gamma_{\nu\lambda}^{\kappa} + \cdots$ .

Looking for static and spherically symmetric electric solutions, we assume a radial electric field $F_{tr}(r)$ and a static spherically-symmetric spacetime:
\begin{equation}
ds^2= u(r)dt^2 - dr^2/f(r) - r^ 2d\Omega^2
\; .
\label{EqStaticSphericalMetric}
\end{equation}

In the simple case of a static radial electric field, it is obvious that $T_t^t=T_r^r$ which yields the simplifying relation $u(r)=f(r)$. There are thus only two independent components of $T_\mu^\nu$, since spherical symmetry results also $T_\theta^\theta=T_\phi^\phi$. These components are still given by the expressions in Eq. \eqref{TmunuSphericalElField-NLED2}. Thus, there are only two algebraically independent Einstein equations, but they are related through the Bianchi identities, so only one of them is needed to obtain the metric function $f(r)$. We naturally choose the first order $(tt)$ equation
\begin{eqnarray}
\frac{1}{r^2 }\frac{d}{dr}r(1-f)&=&\kappa \left(\frac{1}{2}F_{tr}^2+\frac{(2n-1)2^{n} |\gamma|}{2n} F_{tr}^{2n} \right) \nonumber \\
\quad \Rightarrow \quad
\frac{dM}{dr} &=& \frac{\kappa \mathfrak{E}^{2} r^2 }{2} \left( \frac{1}{2}\mathcal{E}^2+\frac{(2n-1)}{2n}\mathcal{E}^{2n} \right) \;,
\label{EinsteinEq00FFn2ndOrder}
\end{eqnarray}
where we defined the mass function as usual by $f(r)=1-2M(r)/r$ and used the dimensionless field variable $\mathcal{E}=F_{tr}/\mathfrak{E}$ (still keeping the additional condition $(-1)^n\gamma >0$).

The field  $F_{tr}$ is also still given by the same expression of flat space - Eq. \eqref{EqFtrSecondOrderGen_n}, which we prefer to rewrite here in a dimensionless form making use of the length scale $\ell = 1/\sqrt{\kappa \mathfrak{E}^{2}}$:
\begin{equation}
\mathcal{E}+\mathcal{E}^{2n-1} = q/\varrho^2
\label{EqFtrSecondOrderGen_nDmlss}
\end{equation}
where $\varrho=r/\ell$ and $q = \kappa \mathfrak{E}\, Q$. We will also need the dimensionless mass function $m=M/\ell$ in order to convert also \eqref{EinsteinEq00FFn2ndOrder} into a dimensionless form.

As in flat space, we do not have at our disposal a simple expression for $\mathcal{E}(\varrho)$, but only for the inverse function $\varrho(\mathcal{E})$. Therefore, we will need to solve for $f(\mathcal{E})$ or
for $m(\mathcal{E})$ in order to get the radial dependence in a parametric form $(f(\mathcal{E}),\varrho(\mathcal{E}))$ or $(m(\mathcal{E}),\varrho(\mathcal{E}))$. The easiest is to solve for $m(\mathcal{E})$ the converted $M$-equation of \eqref{EinsteinEq00FFn2ndOrder}:
\begin{equation}
\frac{dm}{d\mathcal{E}} = \frac{\varrho^{2}(\mathcal{E}) }{2} \left(\frac{1}{2}\mathcal{E}^2+\frac{(2n-1)}{2n} \mathcal{E}^{2n} \right)\frac{d\varrho}{d\mathcal{E}}
\label{EqStaticSphericalMetric2}.
\end{equation}
Integrating this equation we find the following expression in terms of Gauss hypergeometric functions\cite{A+S1965} $F(a,b, c, z)$:
\begin{eqnarray}
m(\mathcal{E})= q^{3/2} \left(\frac{ 5 n+ (6 n-1) \mathcal{E}^{2
   (n-1)}}{12 n \left(1+ \mathcal{E}^{2(n-1)}\right)^{3/2}}\mathcal{E}^{1/2}
 \hspace{4.0cm} \right. \nonumber \\ \left.
+\frac{2 }{3 (2 n-3) \, \mathcal{E}^{n-\frac{3}{2}}}  \,
  F\left(\frac{1}{2},\frac{2 n-3}{4 (n-1)},\frac{6 n-7}{4 (n-1)},
  -\frac{1}{ \mathcal{E}^{2 (n-1)}}\right)  \right)+ m_0
   \label{MassFunctionNLED2ndOrder}
\end{eqnarray}
where $m_0 $ is an integration constant which plays the role of a point mass that may reside at the origin ($r=0$). The total mass of the BH is the sum of this mass together with the field energy which is the same as the one which was calculated in flat space\footnote{Notice the factor $\kappa/(8\pi)$ which multiplies here the field energy of \eqref{TotalEnergy2ndOrderNLED-Analytic}, since the mass function contains a factor of the Newton constant $G$. Recall also the relation $q = \kappa \mathfrak{E}\, Q$. And finally, in order to avoid confusion, the BH masses are designated here with the subscript $BH$. We will not do it in what follows, where there is no risk.}:
\begin{equation}
M_{_{BH}}=\ell\; m_{_{BH}} = \ell \lim_{\mathcal{E}\rightarrow 0}{m(\mathcal{E})} = M_0  + \frac{ 2\kappa Q^{3/2} \mathfrak{E}^{1/2}}{3 \sqrt{\pi } }
  \,\cdot\, \frac{\Gamma\left(\frac{1}{4 (n-1)}\right) \Gamma \left(\frac{6 n-7}{4(n-1)}\right)}{2n-3 }
   \label{BHMassAndM0NLED2ndOrder}
   \end{equation}
   The metric function $f(r)$ is given in parametric form as \vspace{0.05cm}
\begin{eqnarray}
\left\{
\begin{array}{rl}
&f(\mathcal{E})= 1-\frac{\displaystyle{2 }}{ \displaystyle{q^{1/2} }}\left(\mathcal{E}+\mathcal{E}^{2n-1}\right)^{1/2} m(\mathcal{E})  \vspace{0.2cm}\\ 
&r(\mathcal{E})= \ell \cdot \varrho(\mathcal{E}) =  \frac{ \displaystyle{ \ell q^{1/2}}}{\displaystyle{\left(\mathcal{E}+\mathcal{E}^{2n-1}\right)^{1/2}}}  \, \, \, , \\
\end{array} \right.
   \label{MetricFctnNLED2ndOrder}
\end{eqnarray}

 \begin{figure}[t!]
\begin{center}
\includegraphics[width=0.49\textwidth]{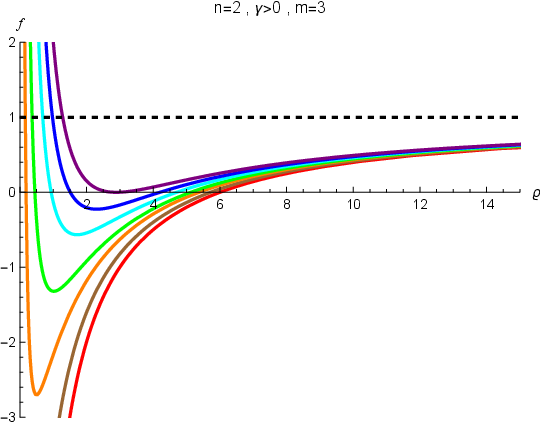}
\includegraphics[width=0.49\textwidth]{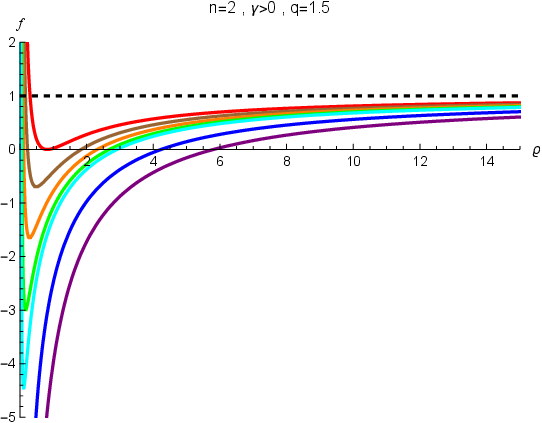}
\caption{\small{Plots of the BH metric function $f(\varrho)$ of  GHE model  with fixed mass and several charges and the other way around. Left: $m=3$  and charges: $q =0,\; 0.25,\; 1.80602,\; 2.6,\; 3.0,\; 3.5,\; 3.9,\; 4.285(\text{EBH})$. The 2 lowest curves with the smallest $q$ values cannot be resolved in this scale and look both red. Right: $q=1.5$ and masses:  $m =0.999(\text{EBH}),\; 1.22,\; 1.4,$ $\; 1.55,\; 1.65\;, 2.25,\; 3.0$. The ``spectroscopic'' color code is used here as in Fig.\ref{FigProfilesNLED2} and in most of the following figures.}}
\label{Fig-f-BH-NLED2nd}
\end{center}
\end{figure}
where $m(\mathcal{E})$ is given by the explicit form of \eqref{MassFunctionNLED2ndOrder}. Fig. \ref{Fig-f-BH-NLED2nd} shows typical $f(r)$ profiles\footnote{Notice that we use the same letter ($f$) for the different functions $f(r)$ and $f(\mathcal{E})$ which represent the same physical quantity, i.e. the metric component. We will keep using this convention throughout this paper.} of electric BHs with fixed mass and several charges and the complementary ones with fixed charge and varying mass. Since the higher $n$ BHs are not significantly different from the $n=2$ ones, it will suffice to use the $n=2$ results to demonstrate all their general characteristics. 

It is obvious that there are two distinct types of solutions characterized by the ratio $Q/M$. For a given $M$, small charge BHs have a Schwarzschild-like (``S-like'') behavior with a single horizon. For higher charges, above a certain critical charge, a second horizon appears and the behavior becomes Reissner-Nordström-like (``RN-like'') with an inner (Cauchy) and outer (event) horizons. The RN-like behavior  extends up to a maximal value of charge, above which the solutions exhibit naked singularities.

Notice especially the soliton-like BH which is ``made of'' a non-trivial configuration of the self-interacting vector field without a point mass in the origin (i.e. $m_0=0$). Still, this solution is singular at the origin as is seen from the electric field and the metric component $f(r)$ or the energy density.

More quantitatively, it is easy to see that this $m_0=0$ differentiates between the two kinds of solutions: The monotonically increasing S-like solutions appear if $m_0 \geq 0 $, while the RN-like ones appear for $m_0 < 0 $. The transition point is defined by the pair $(q,m)$  which satisfies the dimensionless version of \eqref{BHMassAndM0NLED2ndOrder} with $m_0 = M_0 =0 $:
\begin{equation}
m_{_{BH}} =\overline{m}_{field} = \frac{ 2q^{3/2} }{3 \sqrt{\pi } }
  \,\cdot\, \frac{\Gamma\left(\frac{1}{4 (n-1)}\right) \Gamma \left(\frac{6 n-7}{4(n-1)}\right)}{2n-3 }
   \label{BHMassSeparateNLED2ndOrderDmlss} .
\end{equation}
For a fixed charge, this mass is the minimal mass of the S-like solutions. Below this mass RN-like solutions exist down to the minimal mass of the extremal BH (EBH) to be discussed below. Taking a fixed mass, Eq.\eqref{BHMassSeparateNLED2ndOrderDmlss} determines the maximal charge of the S-like solutions. Larger charges correspond to the RN-like BHs up to the maximal charge of the extremal solution. For example, for the profiles with $m=3$ shown in Fig.\ref{Fig-f-BH-NLED2nd} left, the transition from S-like behavior to RN occurs at $q = 1.80602$ for which $\overline{m}_{field} =3$.

A second important point in parameter space is the location of the EBHs. But the charge - mass relation of the EBHs can be obtained only in a parametric form involving the horizon radius $\varrho_h$ or the corresponding (dimensionless) electric field variable $\mathcal{E}_h = \mathcal{E}(\varrho_h)$.

So we turn to present the relation between the BH mass, charge and horizon radius. This relation may be  obtained from the condition $f(\varrho_h)=0$ 
but since we have $f(\varrho)$ only in a parametric form, it will be more practical to get it from the upper equation of \eqref{MetricFctnNLED2ndOrder} and obtain the BH mass as a function of the horizon  field variable, $\mathcal{E}_h$:
\begin{eqnarray}  
m_{_{BH}} = \frac{ \displaystyle{ q^{1/2}}}{2\displaystyle{\left(\mathcal{E}_h+\mathcal{E}_h^{2n-1}\right)^{1/2}}}+
  \frac{2 q^{3/2} }{3 \sqrt{\pi } }
  \,\cdot\, \frac{\Gamma\left(\frac{1}{4 (n-1)}\right) \Gamma \left(\frac{6 n-7}{4(n-1)}\right)}{2n-3} \hspace{5.0cm}    \nonumber   \\
   -q^{3/2} \left[\frac{ 5 n+ (6 n-1) \mathcal{E}_h^{2
   (n-1)}}{12 n \left(1+ \mathcal{E}_h^{2(n-1)}\right)^{3/2}}\mathcal{E}_h^{1/2}
+\frac{2 }{3 (2 n-3) \, \mathcal{E}_h^{n-\frac{3}{2}}}  \,
  F\left(\frac{1}{2},\frac{2 n-3}{4 (n-1)},\frac{6 n-7}{4 (n-1)},
  -\frac{1}{ \mathcal{E}_h^{2 (n-1)}}\right)  \right] 
  \label{BHMassAndHorizonNLED2ndOrderDmlss}.
 \end{eqnarray}
Combining this expression with  $\varrho(\mathcal{E}_h)$ from \eqref{MetricFctnNLED2ndOrder}, we obtain a parametric representation of the $r_h$ dependence of the BH mass. Actually, the result of \eqref{BHMassAndHorizonNLED2ndOrderDmlss} may be written in the following compact form 
\be
m_{_{BH}} =\overline{m}_{field}+\frac{1}{2}\varrho(\mathcal{E}_h)-m_{field}(\mathcal{E}_h)
\label{NLED2MassvsHorizons-compact} ,
\ee
where $\overline{m}_{field}$ is the total field energy given in \eqref{BHMassSeparateNLED2ndOrderDmlss}, $\varrho(\mathcal{E}_h)$ is the general relation from \eqref{MetricFctnNLED2ndOrder} between $\varrho$ and  $\mathcal{E}$  calculated for $\mathcal{E}=\mathcal{E}_h$ and  $m_{field}(\mathcal{E}_h)$ is the horizon value of the field mass function -- the first term of the RHS of \eqref{MassFunctionNLED2ndOrder} which represents the accumulating field energy.

\begin{figure}[t!]
\begin{center}
\includegraphics[width=0.49\textwidth]{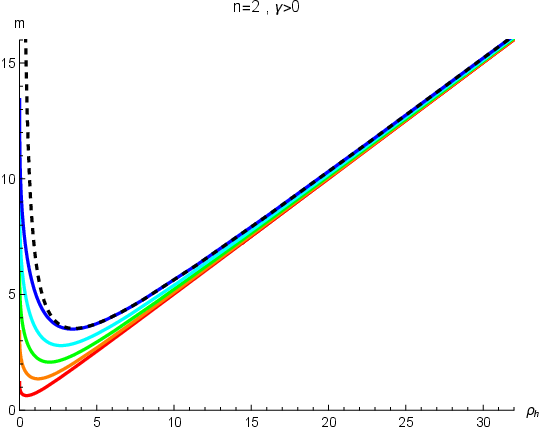}
\includegraphics[width=0.49\textwidth]{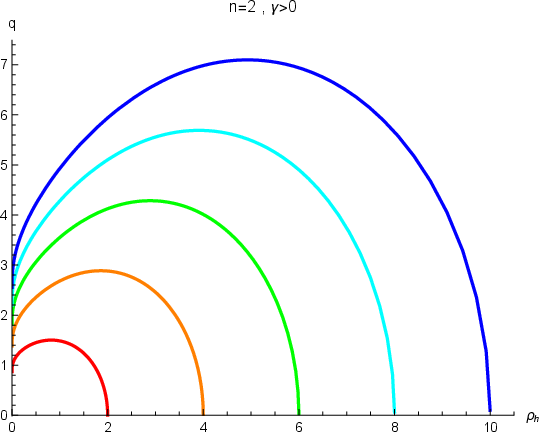}
\caption{\small{Plots of the BH mass (left panel) and charge (right panel) vs the horizon radius $\varrho_h$. The mass curves correspond to the following values of charge: $q =1,\; 2,\; 3,\; 4,\; 5 $. The charge curves correspond to the following values of mass: $m =1,\; 2,\; 3,\; 4,\; 5 $. The dashed line is the RN curve ($\gamma=0$) with $q=5$. Notice the finite mass and charge at $\varrho_h =0$ unlike for RN.} }
\label{Fig-BHmass+ChargeVsrH-NLED2nd}
\end{center}
\end{figure}

The behavior of $m(\varrho_h)$ shares with the well-known RN case its main feature of having a minimum at the extremal BH solution. However, since the RN-like branch of solutions terminates at a finite value of $(q,m)$ related by Eq.\eqref{BHMassSeparateNLED2ndOrderDmlss}, the curve of $m(\varrho_h)$ meets the $\varrho_h = 0$ line at a finite value of mass unlike its RN analogue. At the other end of $\varrho_h$ values, the curve increases monotonically and approaches a linear asymptotic behavior as for RN. These features are obvious from the left panel of Fig. \ref{Fig-BHmass+ChargeVsrH-NLED2nd}, which indicates that the GHE model regularizes the central singularity of the field energy density thus preventing the divergence of the mass function at small radii. The complementary right panel visualizes the charge dependence of these BHs. It may be useful to stress that the physical regions of the 2 plots of  of Fig. \ref{Fig-BHmass+ChargeVsrH-NLED2nd} are limited to the right of the points of minimal mass in the left panel and of the maximal charge in the right panel. the region to the left corresponds to the inner (Cauchy) horizons when they exist. 

Now we can return to the extremal BHs for which the second (extremality) condition $f'(\varrho_h)=0$ should be satisfied in addition to $f(\varrho_h)=0)$ which is solved by Eq. \eqref{BHMassAndHorizonNLED2ndOrderDmlss}. Writing $f(\varrho)$ as
 $f(\varrho)=1-2m(\varrho)/\varrho$ we find that  $f'(\varrho_h)=(1-2m'(\varrho_h))/\varrho_h $, so the extremality condition is $2m'(\varrho_h)=1$. We can rewrite this condition in terms of $\varrho_h$ and  $\mathcal{E}_h$ using Einstein equation \eqref{EinsteinEq00FFn2ndOrder} as:
\begin{equation}
 \varrho_h^{2}\left(\frac{1}{2}\mathcal{E}_h^2+\frac{(2n-1)}{2n} \mathcal{E}_h^{2n} \right)=1
\label{ExtremalityCondGHE}
\end{equation}
and after using the explicit expression for $\varrho_h(\mathcal{E}_h)$ from \eqref{MetricFctnNLED2ndOrder} we obtain  the  extremality condition as the following expression for the electric charge in terms of the horizon field:
\begin{equation}
 q_{_{EBH}}=\frac{2n }{n\mathcal{E}_h+ (2n-1)\mathcal{E}_h^{2n-1}}.
\label{ExtremalBHChargeGHE}
\end{equation}
The associated mass, $m_{_{EBH}}$ is obtained in a straightforward way by substituting Eq.\eqref{ExtremalBHChargeGHE}  into \eqref{BHMassAndHorizonNLED2ndOrderDmlss}.  The explicit expression is quite long and cumbersome and we do not write it down.

\subsection{Thermodynamical Properties of GHE Electrostatic Black Holes}
\label{GHEBHs-Thermodynamics}

\begin{figure}[b!]
		\centering
		\includegraphics[width=0.43\textwidth]{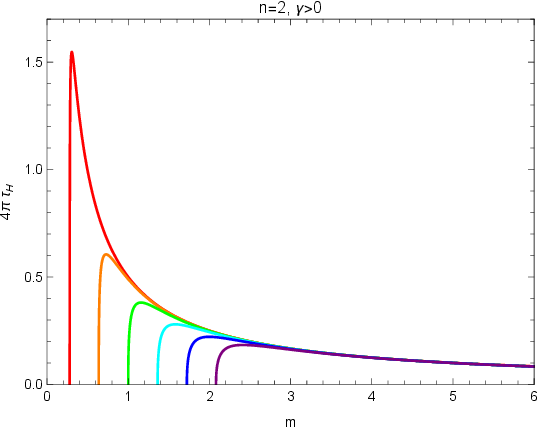} 
        \includegraphics[width=0.5\textwidth]{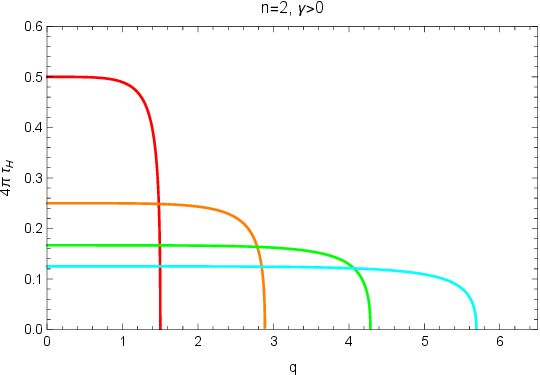}
		\caption{\small The behavior of the rescaled Hawking temperature $\tau_H$ (times $4\pi$) for $n=2$ and $\gamma > 0$. Left: as a function of the BH mass for several values of the charge parameter: $q = 0.5, 1, 1.5, 2, 2.5,3$. Right: as a function of the charge parameter $q$  for several values of the BH mass $m = 1, 2, 3, 4$.}
		\label{fig:Temp2mH+Q-Param}
\end{figure}
The Hawking temperature of a black hole with the static spherically symmetric metric (\ref{EqStaticSphericalMetric}) with $u(r)=f(r)$
 is expressed by ($k_B$ is the Boltzmann constant)
\begin{equation}
T_H=\frac{\hbar}{4 \pi k_B} f^\prime (r)|_{r=r_{h}}
\label{THsphericalEM} .
\end{equation}

Using Eq.(\ref{EinsteinEq00FFn2ndOrder}) for expressing $f^\prime (r_h)$ in terms of the horizon radius and electric field we get the simple result in dimensionless form:
\begin{equation}
\frac{\ell k_B}{\hbar} T_H=\tau_H= \frac{1}{4 \pi}\left[\frac{1}{\varrho_h}-\left(\frac{1}{2}\mathcal{E}_h^2+\frac{2n-1}{2n} \mathcal{E}_h^{2n} \right)\varrho_h \right].
\label{BHTempNLED2ndOrderDmlss}
\end{equation}
where the temperature is rescaled by $\hbar/ k_B\ell$ to yield the dimensionless temperature variable of this system $\tau_H$. 

The dependence on $r_h$ (or $\varrho_h$) of the temperature of a BH with a given charge can be obtained in a parametric form, using the above equation (\ref{BHTempNLED2ndOrderDmlss}) combined with the expression for $\varrho(\mathcal{E})$ from Eq. (\ref{MetricFctnNLED2ndOrder}). The mass dependence can be obtained similarly using the BH mass from Eq. (\ref{BHMassAndHorizonNLED2ndOrderDmlss}) instead of Eq. (\ref{MetricFctnNLED2ndOrder}). We also notice that the extremality condition (\ref{ExtremalityCondGHE}) causes the corresponding BH temperature to vanish in accordance with \eqref{THsphericalEM}. Fig. \ref{fig:Temp2mH+Q-Param} shows in more detail the dependence of the BH temperature on the BH mass and charge. Notice the different (but consistent of course) realizations of the EBH points and the maximal temperatures.  We leave out the $r_h$-dependence, since it does not add much to the other plots.  
\begin{figure}[b!]
		\centering
		\includegraphics[width=0.5\textwidth]{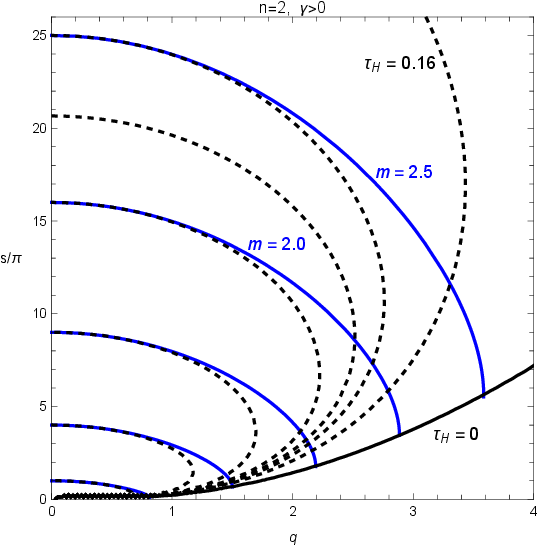}
\caption{\small Curves of the GHE BH rescaled entropy  as a function of charge for various values of mass  (full lines) and  temperature (dashed lines except $T=0$) in the typical case of $n=2,\, \gamma>0$. The  parameters are      
        $m_{_{BH}}=0.5, 1, 1.5, 2, 2.5$, and $4\pi\tau_H = 0, 0.16, 0.2, 0.22, 0.25, 0.33, 0.5, 1.00$. Several numerical values are added in order to indicate the way $m_{_{BH}}$ and $\tau_H$ increase.}
\label{fig:entropyPar}
\end{figure}

The Hawking-Bekenstein entropy of a spherically symmetric black hole or its dimensionless counterpart ($s$) are given by the area law 
\begin{align}
S_{BH} = \frac{k_B}{4l_P^2} A_h \;\;\; \Rightarrow  \;\;\;
s=\frac{S_{BH}}{k_B} \left ( \frac{l_P}{\ell}  \right )^2 = \pi \varrho_h^2
\label{BH-Entropy}
\end{align}
where $A_h=4\pi r_h^2$ is the area of the black hole horizon and $l_P = \sqrt{G_N \hbar}$ is the Planck length. Since it is more practical in this case to use the field variable $\mathcal{E}$ instead of $\varrho$, the way to study the interrelations between the entropy, mass, charge and temperature is to keep using the parametric representations. The appropriate expression for the entropy is evidently 
\begin{equation}
s= \frac{ \displaystyle{\pi q}}{\displaystyle{\mathcal{E}_h+\mathcal{E}_h^{2n-1}}}
\label{BH-EntropyNLED2},
\end{equation}
which is to be used together with Eqs.  \eqref{BHMassAndHorizonNLED2ndOrderDmlss} and \eqref{BHTempNLED2ndOrderDmlss} in order to clarify the thermodynamic aspects of these BHs. Fig.\ref{fig:entropyPar} shows a typical representation of the entropy dependence on the BH mass and charge and also the isotherms in the $(s,q)$ plane. 


The thermodynamic stability of black holes is investigated by the
behaviour of their heat capacity $C$, defined as $G_{N}C=\partial M/\partial T$ (recall that $M$ contains a multiplicative factor of $G_N$). The positive (negative) heat signifies the local thermodynamic stability (instability) of the black holes. The dimensionless version of the heat capacity is given by
\begin{equation}
\frac{l_P^2}{k_B \ell^2} C = c = \frac{\partial m }{\partial \tau_{H}} = \frac{\partial m / \partial \mathcal{E}_h }{\partial \tau_{H} / \partial \mathcal{E}_h}
\label{BH-HeatCapacityNLED2}\, ,
\end{equation}
where  the last term should be used for concrete calculations, since all functions are expressed in terms of the field parameter $\mathcal{E}$. This representation indicates that the heat capacity exhibits a singularity when the Hawking temperature reaches an extreme value. Using the compact equation \eqref{NLED2MassvsHorizons-compact} for the BH mass and the field equation \eqref{EqStaticSphericalMetric2} one finds the following expression for the rescaled heat capacity of the black hole 
\begin{eqnarray}
 \frac{c}{4\pi}=-\frac{q  \left(2n  - n q\mathcal{E}_h+2 n \mathcal{E}_h^{2(n-1)}-(2 n-1) q \mathcal{E}_h^{2 n-1}\right)}
 {2 \left(\mathcal{E}_h+\mathcal{E}_h^{2 n-1}\right) \left(2n -3n q \mathcal{E}_h +2 n \mathcal{E}_h^{2( n-1)} -
 (2n+1) q \mathcal{E}_h^{2 n-1}\right)}
   \label{NLED2HeatCap}.
\end{eqnarray}

\begin{figure}[b!]
		\centering
		\includegraphics[width=0.49\textwidth]{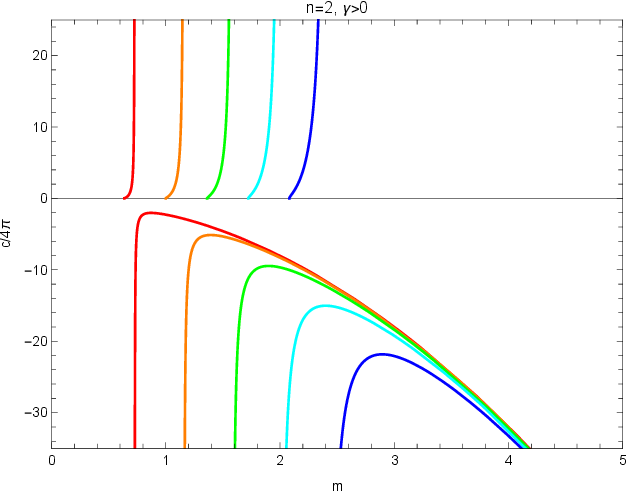} 
        \includegraphics[width=0.49\textwidth]{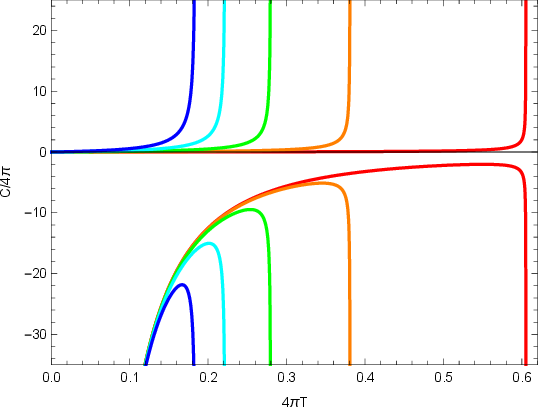}
		\caption{\small The rescaled BH heat capacity as a function of its mass (left) and temperature (right) for  $n = 2$  and $\gamma > 0$. Both for charge values: $q = 1, 1.5, 2, 2.5, 3$. }
		\label{fig:HeatCapacityVSmq}
\end{figure}

Figure \ref{fig:HeatCapacityVSmq} demonstrates the general behavior of the heat capacity as a function of the BH mass, temperature and charge. The most prominent feature is that the  heat capacity gets positive as well as negative values separated by discontinuities at points where the Hawking temperature attains a maximum, signaling the occurrence of phase transitions. Negative heat capacity corresponds to the black hole's unstable state and the early phase of the thermodynamic process, while positive heat capacity corresponds to the stable state of the black hole and the later phase of the thermodynamic process. 

\section{Equivalent first order formulation of NLED}
\label{FirstOrderNLEDversion1}
\setcounter{equation}{0}

It is known \cite{Pellicer+Torrence1969,Plebanski1970,SalazarEtAl1987,Bronnikov2001} that NLED has an equivalent first order formulation where the gauge potential is supplemented by a momentum-like tensor quantity used to perform a Legendre transformation as follows:  Start with the 2nd order NLED Lagrangian $\mathcal{L}^{(2)}(A_{\mu},F_{\mu\nu})$ given in Eq. \eqref{LagNLED2-general}, define covariant ``conjugate momenta'' by $\it{\Pi}^{\mu\nu}=\partial \mathcal{L}/\partial F_{\mu\nu}$ and perform a Legendre transformation which results a ``Hamiltonian'' (or rather a ``Legendrian'') $\mathcal{H}={\it{\Pi}}^{\mu\nu}F_{\mu\nu}-\mathcal{L}^{(2)}$ where the ``velocities'' $ F_{\mu\nu}$ are expressed everywhere in terms of the ``momenta'' $\it{\Pi}^{\mu\nu}$. The field equations now turn into the first order Hamilton-like equations:
\begin{equation}
 F_{\mu\nu}=\frac{\partial \mathcal{H}}{\partial \it{\Pi}^{\mu\nu}} \;\;\;\;\;\;\; , \;\;\;\;\;\;\;
  2\nabla_\mu \it{\Pi}^{\mu\nu}=-\frac{\partial \mathcal{H}}{\partial A_{\mu}}=-J^{\nu}
\label{HamiltonEqsNLED}
\end{equation}
where the second version at the RHS results from the fact that the only $A_{\mu}$-dependence is contained in the standard electromagnetic coupling $J^{\mu} A_{\mu}$ . These field equations may be obtained also by the standard variational procedure\cite{GolsteinBook3} from a first order Lagrangian (density) which depends on the gauge potential, its derivatives (or ``velocities'') and its ``conjugate momenta'' (but not on their derivatives):
\begin{equation}
\mathcal{L}^{(1)}(A_{\lambda},F_{\mu\nu},\it{\Pi}^{\rho\sigma})=\it{\Pi}^{\mu\nu}F_{\mu\nu} - \mathcal{H}(A_{\lambda},\it{\Pi}^{\mu\nu})
\label{LagForHamiltonEqsNLED} .
\end{equation}
We notice that if we define $P^{\mu\nu}=-2{\it{\Pi}}^{\mu\nu}$ we have at hand a second field tensor (which we name ``$P$-field'') in terms of which the source-full field equation will have the same structure as the original source-full Maxwell equation of the linear theory, or even more similarly, to the structure of Maxwell equations in non-linear matter.

Therefore, we may approach NLED theories by starting from a given Legendrian $\mathcal{H}(A_{\lambda},P^{\mu\nu})$ and analyzing its Hamilton equations:
\begin{equation}
 F_{\mu\nu}=-2\frac{\partial \mathcal{H}}{\partial P^{\mu\nu}} \;\;\;\;\;\;\; , \;\;\;\;\;\;\;
 \nabla_\mu P^{\mu\nu}=J^\nu\ .
\label{HamiltonEqsNLEDver2}
\end{equation}
From this point of view the Legendrian $\mathcal{H}$ will be a function of the only two gauge-invariant Lorentz scalars at our disposal, namely the two invariants  $Z=P_{\mu\nu}P^{\mu\nu}$ and $\Omega=P_{\mu\nu}\,^{*}P^{\mu\nu}$ in a clear analogy to $X$ and $\Xi$ defined in the previous section. Thus, from this vantage point, the general NLED theory may be formulated in terms of the general Legendrian (compare \eqref{LagNLED2-general})
\begin{equation}
\mathcal{H} = \frac{1}{4}h(Z, \Omega) + J^{\mu}A_{\mu}\ ,
\label{GeneralLegendrian}
\end{equation}
which yields the following field equations ($h_{_Z}$ and $h_{_\Omega}$   are the partial derivatives of $h(Z, \Omega)$):
\begin{equation}
F^{\mu\nu} = -h_{_Z} P^{\mu\nu} - h_{_\Omega} {^{*}P^{\mu\nu}}\;\;\; , \;\;\;\;\;\;\; \nabla_\mu P^{\mu\nu}=J^\nu\ .
\label{HamiltonEqsNLEDver3}
\end{equation}
The Maxwell limit is obtained for weak fields if the partial derivatives behave like $h_{Z} \rightarrow -1$ and $h_{\Omega} \rightarrow 0$ . Since $F$ is still a closed 2-form, the first equation of \eqref{HamiltonEqsNLEDver3} may be rewritten in a more ``Maxwellian'' form which together with the source-full field equation will form the following set:
\begin{equation}
\nabla_\mu [h_{_Z} \,^{*}P^{\mu\nu} - h_{_\Omega}P^{\mu\nu}]=0 \;\;\; , \;\;\;\;\;\;\; \nabla_\mu P^{\mu\nu}=J^\nu
 \ .
\label{NLEDMaxwellEqs3}
\end{equation}

As is obvious from Eq.\eqref{LagForHamiltonEqsNLED}, these field equations can be also obtained from the variational principle based on the following Lagrangian (defining also for short $Y=P^{\mu\nu}F_{\mu\nu}$):
\begin{equation}
\mathcal{L}_{\rm NLED}^{(1)}= -\frac{1}{4}h(Z,\Omega)-\frac{1}{2}Y -
J^\mu A_\mu\ .
\label{LagNLED1stOrder3}
\end{equation}
This Lagrangian may be also regarded as a generalization to the non-linear regime, of the Palatini formulation of the Maxwell theory discussed in Sec. \ref{FirstOrderMax} . The action in this formulation depends on the gauge potential $A_\mu$ and a field-strength tensor $P_{\mu\nu}$ with no prior assumption about the relation between them. The necessary relation is contained in the source-free field equations which in the non-linear case appear at the left-hand-side (LHS) of Eqs.\eqref{HamiltonEqsNLEDver3} and  \eqref{NLEDMaxwellEqs3}, or more explicitly, in \eqref{F-P-Relations} below.

So, generally speaking, it is a matter of convenience whether to take the 2nd order approach (also known as ``$F$-framework''), or the 1st order (also known as ``$P$-framework''). 

Since these two formulations are considered equivalent, one should study their interrelations and the way to move between them. More technically, starting from a 2nd order formulation defined by a given $f(X,\Xi)$, one has to find the way to construct the 1st order defining function $h(Z,\Omega)$ together with the \textit{invertible} transformation $(X,\Xi)\rightarrow (Z,\Omega)$. The first of the two is easy by comparing the first expression of the ``Legendrian'' $\mathcal{H}$ which leads to Hamilton-like equations \eqref{HamiltonEqsNLED}  with the second, \eqref{GeneralLegendrian}. This gives 
\begin{equation} 
h=2(Xf_{X}+\Xi f_{\Xi})-f(X,\Xi)
\label{1stOrderFromf2ndOrder}.
\end{equation}
However, the function $h$ is expressed here in terms of $X$ and $\Xi$, so we still need to transform them to the other pair $Z$ and $\Omega$. The obvious way to proceed is to compare the field equations of both formulations, \eqref{NLED-FEqs} and  \eqref{HamiltonEqsNLEDver3}. The comparison yields the two ``dual'' relations:
\begin{equation}
F^{\mu\nu} = -h_{_Z} P^{\mu\nu} - h_{_\Omega} {^{*}P^{\mu\nu}} \;\;\; , \;\;\;\;\;\;\;
P^{\mu\nu}= -f_{X}F^{\mu\nu}   -  f_{\Xi}\,^{*}F^{\mu\nu}\ .
\label{F-P-Relations}
\end{equation}
This means of  course that $F$  may be obtained directly from the first equation when the $P$-field is known and vice-versa. It is possible to advance further and obtain the transformation rules between $(X,\Xi)$ and $(Z,\Omega)$, by taking various traces in Eq. \eqref{F-P-Relations} to give after some elementary manipulations:
\begin{equation}
X=Z\left(h_{_Z}^2- h_{_\Omega}^2 \right)+2\Omega h_{_\Omega}h_{_Z}    \;\;\; , \;\;\;
\Xi=\Omega\left(h_{_Z}^2- h_{_\Omega}^2 \right)-2 Z h_{_\Omega}h_{_Z} \ ,
\label{TransfX_Xi-Z_Omega}
\end{equation}
and its inverse
\begin{equation}
Z=X\left(f_X^2- f_{\Xi}^2 \right)+2\Xi f_{\Xi} f_X    \;\;\; , \;\;\;
\Omega=\Xi\left(f_X^2- f_{\Xi}^2 \right)-2X f_{\Xi} f_X\ .
\label{InvTransf-Z_Omega-X_Xi}
\end{equation}
Both these equations give the transformations between $(X,\Xi)$ and $(Z,\Omega)$, but in order to get $h(Z,\Omega)$ from a given $f(X,\Xi)$, the second must be used, serving this way as an implicit relation  which should be inverted to give explicitly $X(Z,\Omega)$ and $\Xi(Z,\Omega)$. Assuming we are able to do this inversion and obtain a well-behaved transformation for a domain which contains the origin, we may directly obtain the sought for explicit expression 
\begin{equation}
h(Z,\Omega)=2[X(Z,\Omega)\Tilde{f}_{X} (Z,\Omega)+\Xi (Z,\Omega) \Tilde{f}_{\Xi} (Z,\Omega) ]-\Tilde{f} (Z,\Omega),
\label{hFromf-Final}
\end{equation}
where the tildes on $\Tilde{f}$, $\Tilde{f}_{X} $ and $\Tilde{f}_{\Xi} $ stress that the change of variables was performed.

We will continue here a little further to present few more useful results which renders this study a more symmetric appearance. 

Next we invert the first equation of \eqref{F-P-Relations} in order to get an independent expression for  $P^{\mu\nu}$ in terms of $F^{\mu\nu}$. The result is
\begin{equation}
P^{\mu\nu}=\frac{h_{_\Omega}{^{*}F^{\mu\nu}}-h_{_Z}F^{\mu\nu}}{h_{_Z}^2+ h_{_\Omega}^2} \ ,
\label{PinTermsF}
\end{equation}
and by comparison with the second equation of \eqref{F-P-Relations}, one finds
\begin{equation}
f_{_X} = \frac{ h_{_Z}}{h_{_Z}^2+ h_{_\Omega}^2} \;\; \ , \;\;\; f_{\Xi} = -\frac{ h_{_\Omega}}{h_{_Z}^2+ h_{_\Omega}^2} \ .
\label{EquivalenceOrders1-2}
\end{equation}

Using now Eqs.\eqref{TransfX_Xi-Z_Omega} and \eqref{EquivalenceOrders1-2} in the Legendre transform \eqref{1stOrderFromf2ndOrder} we get its inverse which yields $f(X, \Xi)$ from a given $h(Z,\Omega)$ in the analogous expression:
\begin{equation} 
f=2(Z h_{Z}+\Omega h_{\Omega})-h(Z,\Omega)
\label{2ndOrderFromf1stOrder},
\end{equation}
which should be supplemented of course by the transformation $(Z,\Omega)\rightarrow (X,\Xi)$  obtained by inverting Eq \eqref{TransfX_Xi-Z_Omega}.

Last, we obtain the following consequence from \eqref{EquivalenceOrders1-2}:
\begin{equation}
(h_{_Z}^2+ h_{_\Omega}^2)(f_{_X}^2+ f_{_\Xi}^2)=1\ .
\label{h-f-QuadraticRelation}
\end{equation}
which gives easily its inverse:
\begin{equation}
h_{_Z} = \frac{ f_{_X}}{f_{_X}^2+ f_{_\Xi}^2} \;\; \ , \;\;\; h_{_\Omega} = -\frac{ f_{_\Xi}}
{f_{_X}^2+ f_{_\Xi}^2} \ ,
\label{InverseEquivalenceOrders1-2}
\end{equation}
supplemented by the transformation $(X,\Xi) \rightarrow (Z,\Omega) $. 

 We should comment that all the above results  are valid only for invertible transformations between the first and second order formalisms. The question of invertibility depends on the nature of the defining functions, either $f(X,\Xi)$ or $h(Z,\Omega)$, and we will return to this point in the next sections.

Finally, as for the 2nd order formulation, we write down the energy-momentum tensor in terms of the relevant quantities for the 1st order formulation defined by Eq. \eqref{LagNLED1stOrder3}:
\be
T_{\mu\nu}= -h_{_Z} P_{\mu\alpha}P_{\nu}^{\ \alpha} - P_{\mu}^{\ \alpha}F_{\nu\alpha}-P_{\nu}^{\ \alpha}F_{\mu\alpha}
-\frac{1}{4}\left(\Omega h_{_\Omega} -h(Z,\Omega) - 2 Y \right)g_{\mu\nu}
\ .
\label{Tmunu1st}
\ee
Using the first equation of \eqref{F-P-Relations}, we may express the energy-momentum tensor in terms of the $P$-frame only:
\be
T_{\mu\nu}= h_{_Z} P_{\mu\alpha}P_{\nu}^{\ \alpha} + h_{_\Omega}\left( P_{\mu}^{\ \alpha}{\;^{*}}P_{\nu\alpha}+P_{\nu}^{\ \alpha}{\;^{*}}P_{\mu\alpha}\right) -\frac{1}{4}\left(3\Omega h_{_\Omega}+2Zh_{_Z} -h(Z,\Omega)  \right)g_{\mu\nu}
\ .
\label{Tmunu1st-InTermsP}
\ee


\section{ Polynomial Electrodynamics in First Order Formalism}
\label{FirstOrder-Poly}
\setcounter{equation}{0}

A natural Lagrangian in the 1st order formalism which is simple enough to study is the  Polynomial  Electrodynamics in First Order Formalism (PEDFOF) defined by the following $h$-function
\begin{equation}
 h_{_{\rm PEDFOF}}(Z,\Omega)= -Z+\frac{2\gamma}{n} Z^n -\alpha \Omega + \frac{2\beta}{n} \Omega^n
\label{hPEDFOF-n},
\end{equation}
and the corresponding 1st order Lagrangian:
\begin{equation}
 \mathcal{L}_{_{\rm PEDFOF}} = \frac{1}{4}Z-\frac{\gamma}{2n} Z^n +\frac{\alpha}{4} \Omega -\frac{\beta}{2n} \Omega^n -\frac{1}{2}Y - J^\mu A_\mu\
\label{LagPEDFOF-n}.
\end{equation}
Notice that the term  linear in $\Omega$ induces a non-trivial dynamical contribution, since  unlike $\Xi$, it is not a total divergence. Taking all the free coefficients ($\alpha$, $\beta$, $\gamma$) to zero gives back the Maxwell theory in 1st order formalism.

The explicit form of the field equations of this Lagrangian will be
\bea
\nabla_\mu P^{\mu\nu}=J^\nu   \hspace{7.2cm} \\
F^{\mu\nu} = (1-2\gamma Z^{n-1}) P^{\mu\nu} + (\alpha-2\beta \Omega^{n-1}) {^{*}P^{\mu\nu}}  \;\;\; \Rightarrow \;\;\;\;\;\;\; \nonumber  \\
\nabla_\mu [(1-2\gamma Z^{n-1}) \,^{*}P^{\mu\nu} + (\alpha-2\beta \Omega^{n-1})P^{\mu\nu}]=0\ .
\label{PEDFOFFEqs}
\eea
The energy-momentum tensor of the vector field of this theory will be expressed in terms of the $P$-field only (see \eqref{Tmunu1st-InTermsP}):
\bea
T_{\mu\nu}=-\left(1- 2\gamma  Z^{n-1} \right)P_{\mu}^{\ \alpha}P_{\nu\alpha} - \left(\alpha- 2\beta  \Omega^{n-1} \right)\left( P_{\mu}^{\ \alpha}{\;^{*}}P_{\nu\alpha}+P_{\nu}^{\ \alpha}{\;^{*}}P_{\mu\alpha}\right)+ \;\;\;\;\;\;\;\; \nonumber \\
\frac{1}{4}\left( Z - \frac{2(2n-1)\gamma}{n}  Z^n + 2\alpha\Omega  - \frac{2(3n-1)\beta}{n} \Omega^n\right)g_{\mu\nu}
 \ .
\label{TmunuPEDFOF}
\eea

\subsection{Flat Space PEDFOF Spherically-Symmetric Electrostatic Solutions}
\label{FOPNLEDSol-Flat}

In order to study purely electrostatic fields of this model, it is enough 
to focus our attention in the  simple PEDFOF Lagrangian where the nonlinearity originates from the $Z^n$ term only while $\alpha$ and $\beta$ vanish. More explicitly, we write the Lagrangian of the PEDFOF ``$Z^n$ model'' as
\begin{equation}
\mathcal{L}_{Z^n}^{(1)}=\frac{1}{4}P_{\mu\nu}P^{\mu\nu}-
\frac{1}{2}P_{\mu\nu}F^{\mu\nu}-\frac{\gamma}{2n}(P_{\mu\nu}P^{\mu\nu})^n  -J^\mu A_\mu \;.
\label{LbM1A} 
\end{equation}
Variation with respect to $A_\mu$ gives again the source-full linear Maxwell equations~\eqref{nonhom} - now written for $P_{\mu\nu}$. Variation with respect to $P_{\mu\nu}$ will produce a modified relation between $P_{\mu\nu}$ and $F_{\mu\nu}$:
\begin{equation}
\left(1-2\gamma Z^{n-1}\right) P_{\mu\nu}=F_{\mu\nu} \ .
\label{LbM1A-Sourceless1}
\end{equation}
Thus $P$ is not a closed 2-form any more, but $F_{\mu\nu}$ is. So we may write all the  nonlinear Maxwell equations in terms of $P_{\mu\nu}$ only, as
\begin{equation}
\nabla_\mu P^{\mu\nu}=J^{\nu} \;\;\; , \;\;\;  \nabla_\mu [\left(1-2\gamma Z^{n-1}\right)\,^{*}P^{\mu\nu}]=0 \ .
\label{LbM1A-Sourcefull+Sourceless2}
\end{equation}

The energy-momentum tensor of the vector field of model~\eqref{LbM1A} is
\begin{equation}
T_{\mu\nu}=(1-2\gamma  Z^{n-1})P_{\mu\alpha}P_{\nu}^{\ \alpha}-P_{\mu\alpha}F_{\nu}^{\ \alpha}-P_{\nu\alpha}F_{\mu}^{\ \alpha}-g_{\mu\nu}\mathcal{L}_{Z^n}^{(1)} \ .
\label{EnergyDensity1stOrderNLED-bM1A}
\end{equation}
Using~\eqref{LbM1A-Sourceless1} this can be written solely in terms of $P_{\mu\nu}$:
 \begin{equation}
T_{\mu\nu}=-\left(1-2\gamma Z^{n-1}\right) P_{\mu\alpha}P_{\nu}^{\ \alpha} +\frac{g_{\mu\nu}}{4}\left(Z - \frac{2(2n-1)\gamma}{n} Z^n  \right)\ .
\label{EnergyDensityVer21stOrderNLED-bM1A}
\end{equation}

In this model the field around a point charge is very easy to obtain since the field strength $P_{\mu\nu}$ satisfies the ordinary Maxwell equations -- see~\eqref{LbM1A-Sourcefull+Sourceless2}. However, now one has to be careful to distinguish between the field strength and the tensor $F_{\mu\nu}$ which is related to $P_{\mu\nu}$ by Eq. \eqref{LbM1A-Sourceless1}. Assuming that the only non-vanishing components of $P_{\mu\nu}$ are $P_{rt}(r)=-P_{tr}(r)$, we find immediately that
\be
P_{tr}=\frac{Q}{r^2} \;\;\;\; , \;\;\;\; F_{tr}=\frac{Q}{r^2}+(-2)^n\gamma \left(\frac{Q}{r^2}\right)^{2n-1}
\label{ElFields-bM1A}
\ee
Conventionally, point particles couple to $A_\mu$ and they ``feel'' the force as $F_{\mu\nu}$ and not as $P_{\mu\nu}$. So we interpret $F_{\mu\nu}$ as the physical field that determines the motion of a test particle in this family of theories.

If we now inspect the energy density of a point charge using \eqref{EnergyDensityVer21stOrderNLED-bM1A}, we come immediately to the conclusion that the field behavior near the origin causes a strongly diverging energy density with no finite total field energy for all values of $n$.

 \begin{figure}[t!]
\begin{center}
\includegraphics[width=0.49\textwidth]{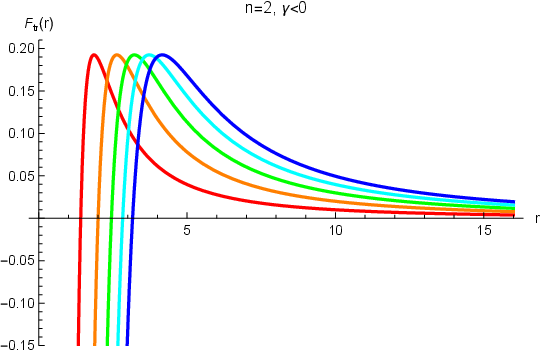}
\includegraphics[width=0.49\textwidth]{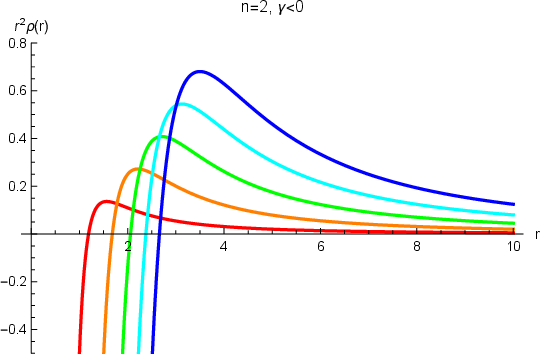}
\caption{\small{Plots of $F_{tr}(r)$ and $r^2\rho(r)$  for $n=2$ and $\gamma<0$ point charge for  several values of $Q=\;1\;,2,\;3,\;4,\;5$  in arbitrary units. The colors correspond to the charge in a ``spectroscopic order''.}}
\label{FigProfilesNLED_Zn2GammaPos}
\end{center}
\end{figure}
Still, it is of interest to write explicitly the components of the energy-momentum tensor for a point charge, or rather for the more general case of a spherically-symmetric electric field:
\bea
T_t^t=T_r^r=  \frac{1}{2}P_{tr}^2+\frac{(-2)^{n} \gamma}{2n} P_{tr}^{2n} \nonumber \\
T_{\theta}^{\theta}=T_{\phi}^{\phi}=-\frac{1}{2}P_{tr}^2-\frac{(2n-1)(-2)^{n} \gamma}{2n} P_{tr}^{2n}
\label{TmunuSphericalElField-bM1A}
\eea 

 \begin{figure}[b!]
\begin{center}
\includegraphics[width=0.49\textwidth]{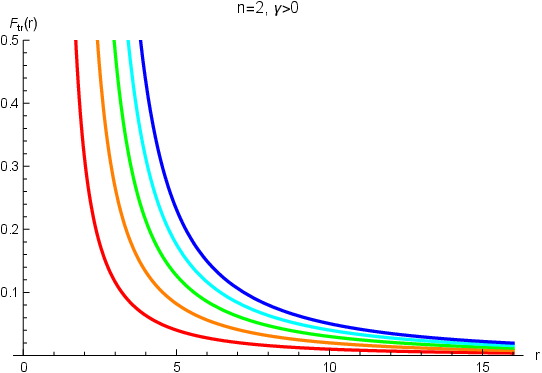}
\includegraphics[width=0.49\textwidth]{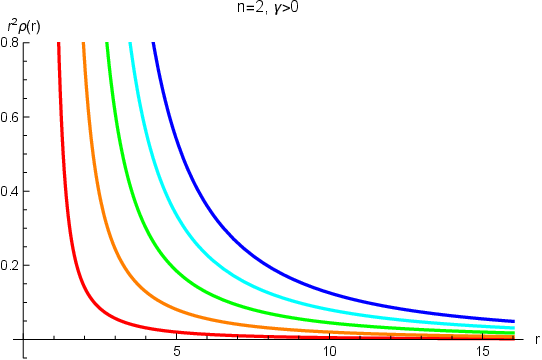}
\caption{\small{Plots of $F_{tr}(r)$ and $r^2\rho(r)$  for $n=2$ and $\gamma>0$ point charge for  several values of $Q=\;1\;,2,\;3,\;4,\;5$ in arbitrary units.}}
\label{FigProfilesNLED_Zn2GammaNeg}
\end{center}
\end{figure}

From this result we see that the energy density $\rho=T_t^t$ of a point charge is positive definite only if we impose $\epsilon=(-1)^n sign(\gamma)=+1$. In the other case ($\epsilon = -1$), the second term of the energy density in \eqref{TmunuSphericalElField-bM1A} will dominate near the origin and will turn the sum to be negative. The same conditions also differentiate between monotonic and non-monotonic behavior of the field $F_{tr}$ of a point charge. Figs \ref{FigProfilesNLED_Zn2GammaPos} and \ref{FigProfilesNLED_Zn2GammaNeg} present the energy density and the field $F_{tr}$ in both cases. 


We can simplify the notation considerably and make the physics more transparent if we notice that we can express the parameter $\gamma$ in terms of 
an electric field parameter $\mathfrak{E}$ by $|\gamma|=1/2^n\mathfrak{E}^{2(n-1)}$ keeping in mind that $\epsilon=\pm 1$ according to the correlation between the parity of $n$ and the sign of $\gamma$: if $(-1)^n$ has the same sign of $\gamma$, $\epsilon = +1$ while otherwise $\epsilon = -1$. Additionally, $\epsilon = 0$  corresponds to the linear theory. Using this choice we can easily give the field equation a dimensionless form absorbing also  $\gamma$ into the electric fields by an appropriate rescaling. Moreover, we can also define a length parameter by $\ell=1/\sqrt{\kappa \mathfrak{E}^2}$ where $\kappa=8\pi G$ is the rescaled Newton's constant, such rendering Einstein equations dimensionless as well. From now on, we will discard the $\epsilon = -1$ solutions due to their non-physical characteristics. We will still keep $\epsilon$ in the equations in order to get control on the limiting behavior to the linear Maxwell theory. 

\subsection{Self-Gravitating PEDFOF Spherically-Symmetric Electrostatic Solutions}
\label{FOPNLEDSol-Grav}
Minimal coupling of gravity to the $Z^n$ model of Eq. \eqref{LbM1A} is described by:
\begin{equation}
\mathcal{L}=\frac{1}{2\kappa}R+\frac{1}{4}P_{\mu\nu}P^{\mu\nu}-\frac{1}{2}P_{\mu\nu}F^{\mu\nu}-
\frac{\gamma}{2n}(P_{\mu\nu}P^{\mu\nu})^n \; ,
\label{EqSelf-Grav-NLED-Z_n}
\end{equation}
which produces Einstein's equations with the energy-momentum tensor given in Eq.\eqref{EnergyDensityVer21stOrderNLED-bM1A}. Since gravity is minimally coupled, it may be treated equivalently by Palatini or metric formalism.

Looking for static and spherically symmetric electric solutions, we assume a radial electric field and a metric tensor as in Eq. \eqref{EqStaticSphericalMetric}. Also, as in Sec.\ref{GHESol-Grav} we have the same simplifications $T_0^0=T_1^1$ and $T_2^2=T_3^3$ which yield $u(r)=f(r)$ and leave only one metric function and   
 two independent components of $T_\mu^\nu$ which are still given by the expressions in Eq. \eqref{TmunuSphericalElField-bM1A}.

The field tensor $P_{01}$ is still given by the same expression of flat space - Eq. \eqref{ElFields-bM1A}, so the single Einstein equation which is required to obtain the metric function $f(r)$ is
\begin{equation}
\frac{1}{r^2 }\frac{d}{dr}r(1-f)=\kappa \left(\frac{Q^2}{2r^4} + \frac{(-2)^{n}\gamma}{2n}\frac{Q^{2n}}{r^{4n}} \right)
\quad \Rightarrow \quad \frac{dm}{d\varrho} = \frac{1}{2} \left(\frac{q^2}{2\varrho^2} + \frac{\epsilon}{2n} \frac{q^{2n}}{\varrho^{2(2n-1)}} \right) \;,
\label{EinsteinEq00Zn}
\end{equation}
where we use the dimensionless radial variable $\varrho=r/\ell=\sqrt{\kappa \mathfrak{E}^2} r$, the dimensionless charge parameter $q=\kappa \mathfrak{E}Q$ and 
defined the mass function as usual by $f(r)=1-2M(r)/r=1-2m(\varrho)/\varrho$. The total mass is given as usual by $M=M(\infty)$ and correspondingly $m=m(\infty)$. Either way the Einstein equation is easily integrated to give the metric component and mass function as:
\be
f(\varrho)=1-\frac{2m}{\varrho}+\frac{q^2}{2\varrho^2}+\frac{\epsilon}{2n(4n-3)} \frac{q^{2n}}{\varrho^{2(2n-1)}}\;\; ; \;\;\;
m(\varrho) = m-\frac{q^2}{4\varrho}-\frac{\epsilon}{4n(4n-3)} \frac{q^{2n}}{\varrho^{4n-3}}
\label{g00_FFn}.
\ee

 \begin{figure}[t!]
\begin{center}
\includegraphics[width=0.49\textwidth]{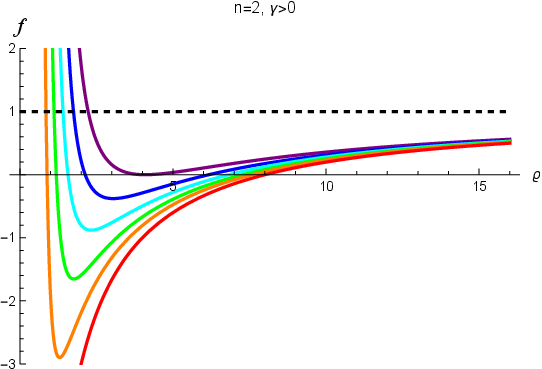}
\includegraphics[width=0.49\textwidth]{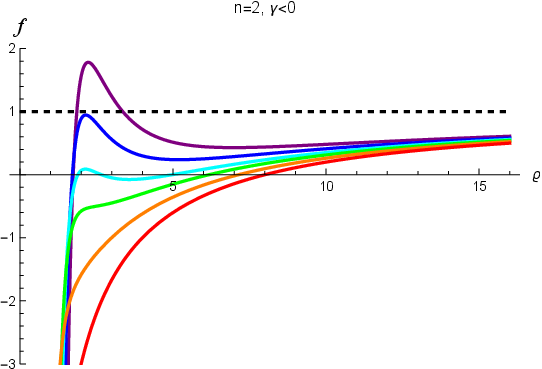}
\caption{\small{Plots of the BH metric function $f(\varrho)$  of the PEDFOF $Z^2$ model with both signs of $\gamma$ with fixed mass and  several values of charge. Left: $\gamma>0$ , $m=4$; $q=0,\;2.5,\;3.25,\;4.0,\;4.75,\;5.624(\text{EBH})$; Right:  $\gamma<0$, $m=4$ ; $q=0,\;3.5,\;4.75,\;5.5,\;6.5,\;7.5$. Notice the 3-horizon profile for $\gamma<0$}}
\label{FigMetricProfilesNLED_FFn2GammaPosNeg}
\end{center}
\end{figure}

Figure \ref{FigMetricProfilesNLED_FFn2GammaPosNeg} shows typical behavior of the metric component $f(r)$ for the two different cases  $\epsilon=(-1)^n sign(\gamma)=-1$ and  $\epsilon=+1$ by plots for $n=2$ with both signs of $\gamma$. Note that the plots of the corresponding electric field $F_{tr}(r)$ and the energy density $\rho(r)$ are still given by those of Figs. \ref{FigProfilesNLED_Zn2GammaPos} and \ref{FigProfilesNLED_Zn2GammaNeg}.

We notice that the general behavior of $f(r)$ is significantly different in both cases for all values of $n$: for $\epsilon>0$ (see left plot of Figure \ref{FigMetricProfilesNLED_FFn2GammaPosNeg}) it is always Reissner-Nordstrom-like with two horizons that are getting closer to each other as $Q$ increases and merge at its critical value at the EBH. Unlike the GHE BHs studied at Sec. \ref{GHE}, the $\epsilon>0$ family does not contain single-horizon S-like BHs. 

For $\epsilon<0$ (right plot of Figure \ref{FigMetricProfilesNLED_FFn2GammaPosNeg}), the behavior starts for small charges as Schwarzschild-like $f(r)$  
monotonically increasing to the asymptotic value of 1 and having a single horizon. However, beyond an intermediate region (see below), above a certain critical charge, $f(r)$ develops one maximum and one minimum before increasing asymptotically to  $f(r)\rightarrow \infty$, still having a single horizon. For intermediate values of charge, it may occur that the local maximum and minimum of $f(r)$  have opposite signs, thus inducing a pair of additional horizons. This intermediate region of solutions is therefore characterized  by the existence of 3-horizons. So, we get to the somewhat surprising conclusion that all these $\epsilon<0$ BH solutions which have negative energy density around the origin, still have attractive gravitational field for large $r$. This result may be in favor of keeping this kind of solutions in the present work, yet we prefer to postpone that to a future study, and to take here the more conservative approach of concentrating in self-gravitating objects with positive definite energy density. So, in the rest of this section we will limit ourselves to $\epsilon\geq 0$ only, $\epsilon = 0$ allowing to include the linear Maxwell theory in the analysis for comparison.      
 \begin{figure}[b!]
\begin{center}
\includegraphics[width=0.40\textwidth]{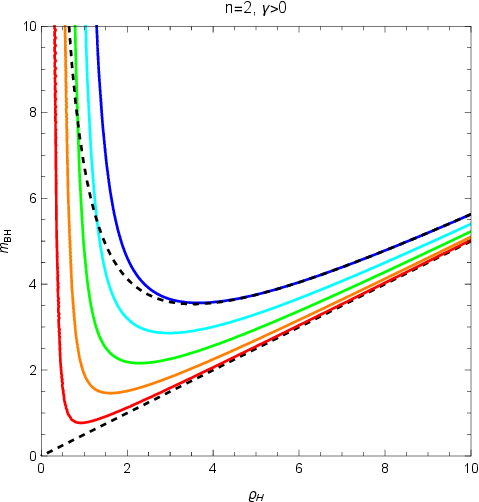}
\includegraphics[width=0.475\textwidth]{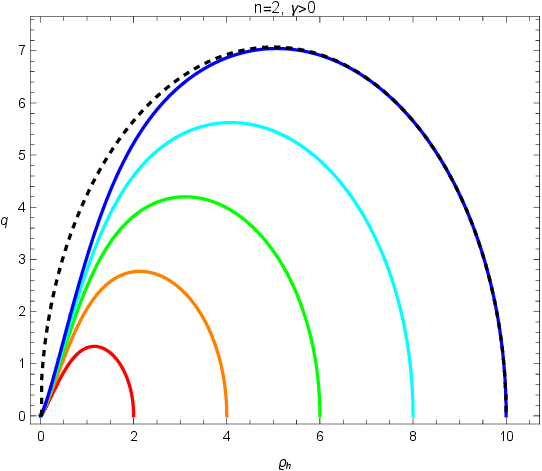}
\caption{\small{Plots of the BH mass vs. the dimensionless horizon size $\varrho_h$ for several values of charge $q = 1, 2, 3, 4, 5$ (left) and charge vs. $\varrho_h$ for several values of mass $m_{_{BH}} = 1, 2, 3, 4, 5$ (right). The dashed black lines correspond to the RN solution (upper curves on both), and to  Schwarzschild solutions (lower left). On the right, Schwarzschild solutions lie along the horizontal axis. }}
\label{FigBHMass+ChargevsrHn2GamPos}
\end{center}
\end{figure}

Next we turn to the relation between the horizon radius and the physical parameters which characterize the black holes, i.e the mass and charge as well as the parameter $n$ while assuming as just mentioned $\epsilon\geq 0$.  

The relation between all these quantities can be obtained by substituting $f(\varrho_h)=0$ in \eqref{g00_FFn} from which we can obtain an expression for $ m_{_{BH}}(n,\epsilon,q,\varrho_h)$:
\be
 m_{_{BH}}(n,\epsilon,q,\varrho_h)=\frac{\varrho_h}{2}+\frac{q^2}{4\varrho_h}+\frac{\epsilon}{4n(4n-3)} \frac{q^{2n}}{\varrho_h^{4n-3}}
\label{MassvsHorizons_FFn} .
\ee
As a trivial check we notice that for $\epsilon =0$ we get the RN result, while if we assume $q=0$, the Schwarzschild relation is obtained.

It is easy to see that the mass function has a single minimum
while $m(n,\epsilon,q,\varrho_h)$ increases indefinitely in both directions. A plot of the mass function as a function of $\varrho_h$ for fixed $n$ and $q$ is shown in Fig.\ref{FigBHMass+ChargevsrHn2GamPos} (left). Its right panel shows the $\varrho_h$-dependence of the charge for a fixed mass (and $n$). Varying $n$ does not change much the general behavior of the solutions.

Notice the usual RN feature that  for fixed charge the mass is bounded from below, while for fixed mass the charge is bounded from above. These features result from the existence of the EBH solutions. Unlike the GHE solutions, in this case it is possible to obtain an explicit relation between the horizon size and EBH charge which is written simply as:
\be
\frac{q^2}{2\varrho_h^{2}}\ + \frac{\epsilon q^{2n}}{2n \varrho_h^{2(2n-1)}} = 1
\label{EBH-Horizon-Charge} .
\ee
Taking $\epsilon=0$ gives the RN result $\varrho_h=q/\sqrt{2}$. Note also that \eqref{EBH-Horizon-Charge} is consistent with the extremality condition $2m'(\varrho_h)=1$ discussed above Eq. \eqref{ExtremalityCondGHE}. See also \eqref{EinsteinEq00Zn}.

It is straightforward to get also an analytic implicit relation between $\varrho_h$ and the EBH mass, but it is quite cumbersome and we give up its presentation.

\subsection{Thermodynamical properties}
\label{PEDFOFBH-Thermo}
In this subsection, we study thermodynamic properties of the black holes presented above, like temperature and heat capacity.
\begin{figure}[b!]
		\centering
		\includegraphics[width=0.45\textwidth]{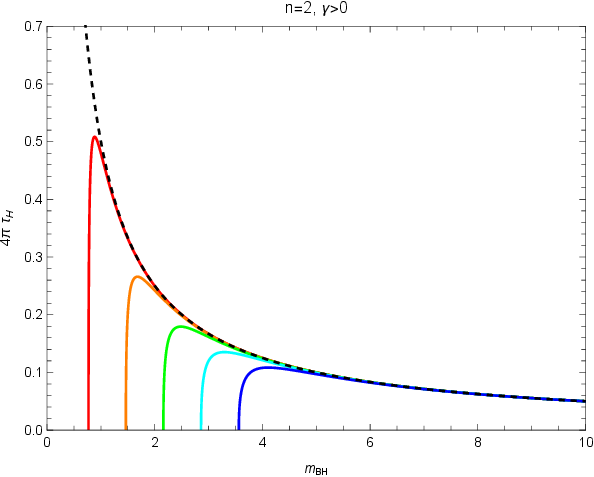} 
        \includegraphics[width=0.53\textwidth]{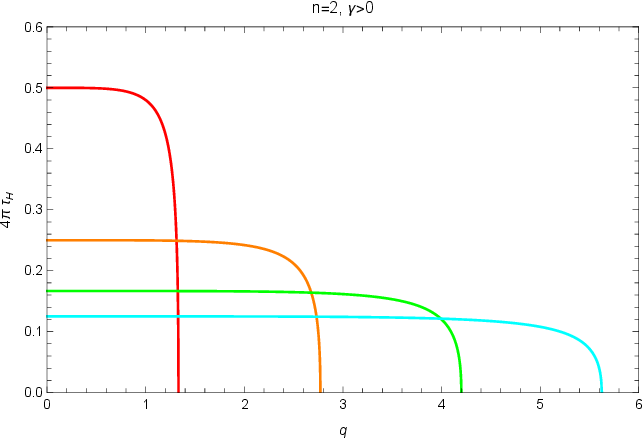}
		\caption{\small 
        The behavior of the rescaled Hawking temperature for $(n=2,\gamma > 0)$ PEDFOF BHs. Left: as a function of the BH mass for $q =1, 1.5, 2, 2.5, 3$. The Schwarzschild BH corresponds to the black dashed line of $q=0$. Right: as a function of the charge parameter $q$  for several values of the mass $m = 1, 2, 3, 4$.}
      		\label{fig:PEDFOFTemp-m+q}
	\end{figure}
   
The Hawking temperature of a black hole with the static spherically symmetric metric (\ref{EqStaticSphericalMetric}) with $u(r)=f(r)$
 is expressed by Eq.(\ref{THsphericalEM})
which gives the following expression for the Hawking temperature using the metric (\ref{g00_FFn}):
\begin{equation}
 \frac{k_B \ell}{\hbar} T_H =\tau_H = \frac{1}{4 \pi} \left ( \frac{1}{\varrho_h} -\frac{q^2}{2 \varrho_h^3} - \frac{\epsilon q^{2 n}}{2n \varrho_h^{4n-1}} \right ) \,
 \label{PEDFOFn2TH}.
\end{equation}
It is trivial to verify that the limit  $q\rightarrow 0$ yields the temperature of the Schwarzschild black holes: $\tau_H= 1/4\pi\varrho_h$ and by taking $\epsilon\rightarrow 0$ one obtains the RN result. The EBH condition \eqref{EBH-Horizon-Charge} guarantees also $\tau_H = 0$ as should be.  

Fig.\ref{fig:PEDFOFTemp-m+q} depicts the way the  PEDFOF BH temperature depends on the BH mass and charge of the $n=2$ family of solutions which is typical and represent the behavior for all $n>2$. The behavior is quite similar to that of the 2nd order GHE BHs of Sec. \ref{GHE} and even more similar to the RN behavior of the linear Maxwell-Einstein system. 
 Looking at Eq.\eqref{PEDFOFn2TH} it is easy to see that the NLED last term modifies the temperature behavior, which for $\epsilon> 0$ and $n\geq 2$ cannot ``spoil'' neither the general characteristics of this relation, nor the behavior at the two ends:  near the EBH point and at large $\varrho_h$ (or $m$). 
  \begin{figure}[t!]
		\centering
		\includegraphics[width=0.5\textwidth]{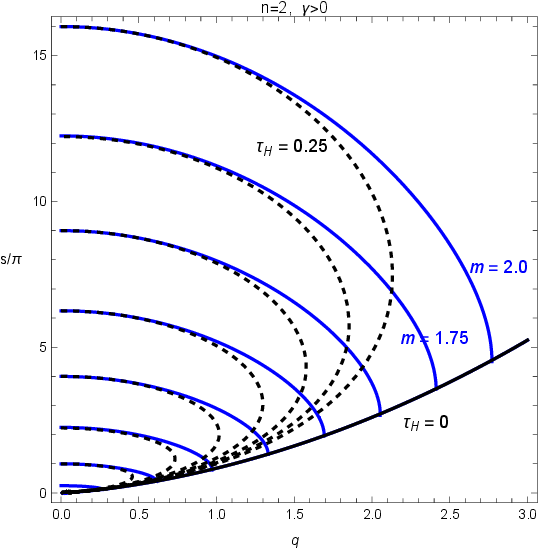}
		\caption{ \small Curves of the  BH rescaled entropy  as a function of charge for various values of mass  (full lines) and  temperature (dashed lines except $T=0$) in the typical case of $n=2,\, \gamma>0$. The  parameters are      
        $m_{_{BH}}=0.25, 0.50, 0.75, 1.00, 1.25, 1.50, 1.75, 2.00$, and $4\pi\tau_H = 0, 0.25, 0.286, 0.333, 0.40, 0.50, 0.67, 1.00$. Several numerical values are added in order to indicate the way $m_{_{BH}}$ and $\tau_H$ increase.}
\label{fig:entropy}
\end{figure}

The Hawking-Bekenstein entropy of a spherically symmetric black hole is given by the area law given already by Eq. \eqref{BH-Entropy}: $s= \pi \varrho_h^2$. Since in this case we use $\varrho$ directly as the radial variable, it is easier to get the relations between $S$, $T$, $M$ and $Q$  than in the previous  GHE case. The main features are seen in  Fig.\ref{fig:entropy} which presents the lines of constant mass and constant temperature (isotherms) in the $q$-$s$ plane. 

Moreover, to check the thermodynamic stability of black holes, the
behaviour of specific heat capacity of the black holes is calculated. The positive (negative) specific heat
signifies the local thermodynamic stability (instability) of the black holes. In the present PEDFOF case we have at our disposal the BH mass as a function of the horizon radius, but not explicitly of the temperature. Therefore, the heat capacity is calculated by an expression analogous to 
\eqref{BH-HeatCapacityNLED2}:
\begin{equation}
\frac{l_P^2}{k_B \ell^2} C = c = \frac{\partial m }{\partial \tau_{H}} = \frac{\partial m / \partial  \varrho_h }{\partial \tau_{H} / \partial  \varrho_h}
\label{BH-HeatCapacityPEDFOF}\, ,
\end{equation}
which indicates that here too the heat capacity exhibits a singularity when the Hawking temperature reaches an extreme value. Using in  \eqref{BH-HeatCapacityPEDFOF} the results \eqref{MassvsHorizons_FFn}  and \eqref{PEDFOFn2TH} for the mass and temperature, one finds the following expression for the heat capacity of the PEDFOF black holes as a function of $\varrho_h$. 
\begin{equation} 
\frac{c} { 4 \pi}= \frac{-\epsilon q^{2n} \varrho_h^6 + n \varrho_h^{4n+2} (2\varrho_h^2 -q^2)}{2\epsilon (4n-1) q^{2n} \varrho_h^4 + 2n \varrho_h^{4n} (3q^2 - 2\varrho_h^2)}\, .
   \label{NLED1PEDFOFHeatCap}
\end{equation}

\begin{figure}[t!]
		\centering
        \includegraphics[width=0.49\textwidth]{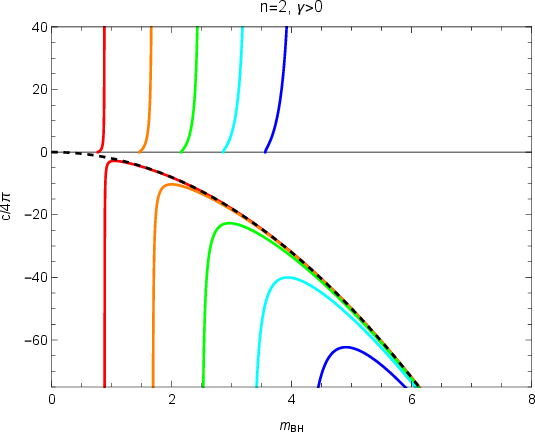}
        \includegraphics[width=0.49\textwidth]{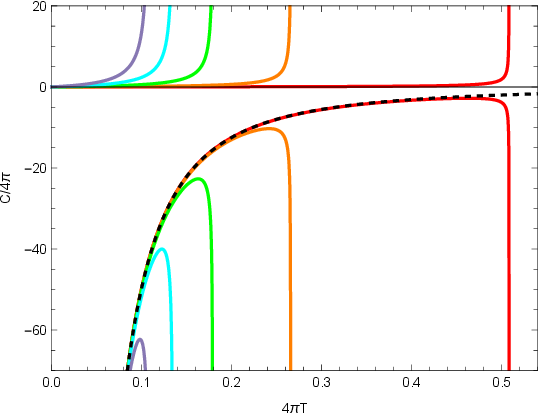}
		\caption{\small{ The rescaled BH heat capacity as a function of its mass (left) and temperature (right) for  $n = 2$  and $\gamma > 0$. Both for charge values: $q = 0, 1, 2, 3, 4, 5$. }}
		\label{fig:HeatHormBH}
	\end{figure}
    
Figure \ref{fig:HeatHormBH} illustrates the general behavior of the heat capacity as a function of the BH mass, temperature and charge. In particular, it shows that the heat capacity diverges at points where the Hawking temperature reaches its peak, signaling phase transitions. The region includes both positive and negative heat capacities, separated by discontinuities. Negative heat capacity corresponds to the black hole's unstable state and the early phase of the thermodynamic process, while positive heat capacity corresponds to the stable state of the black hole and the later phase of the thermodynamic process.

\subsection{PEDFOF Counterpart in Second Order Formulation}
\label{PEDFOF2ndOrder}

Since we discuss in this report NLED in 1st and 2nd order formalism,  a natural question that arises, is what is the 2nd order counterpart of the PEDFOF model presented above (see \eqref{LagPEDFOF-n}). In concrete terms, we look for the 2nd order defining function $f(X,\Xi)$ which corresponds to the 1st order defining function $h(Z,\Omega)$ of \eqref{hPEDFOF-n} and gives the same field equations.

It is known \cite{RuffiniWuXue2013,Breton+Lopez2021} that for weak fields, the functions $f(X,\Xi)$ and $h(Z,\Omega)$ have similar behavior, but this is not sufficient when we deal with solutions featuring strong fields. So we make use of the method  for transforming one to the other which is presented in Sec. \ref{FirstOrderNLEDversion1}. All we have to do is to use the inverse Legendre transform of Eq \eqref{2ndOrderFromf1stOrder} together with the transformation $(Z,\Omega)\rightarrow (X,\Xi)$  which is contained in Eq \eqref{TransfX_Xi-Z_Omega}. For clarity of the demonstration, we will simplify further to the ``$Z^n$ model'' defined by Eq.\eqref{LbM1A} which we used to obtain pure electrostatic solutions (although it allows magnetic solutions as well). For this model we obtain the simple expressions 
\be
f=2Z h_{_{Z}} - h(Z)  \; ,\;\;\; X=Z\,h_{_{Z}}^2
\label{hPEDFOFto2ndOrder}.
\ee
or more explicitly 
\be
f(n,\gamma,Z)=-Z+2\left(2-\frac{1}{n}\right)\gamma Z^n  \; ,\;\;\; X(n,\gamma,Z)=Z(1-2\gamma Z^{n-1})^2
\label{Znto2ndOrder2}.
\ee
A simple inspection reveals that the transformation $X(n,\gamma,Z)$ is not always invertible on the whole real axis, and at any rate, this cannot be done in a closed analytic form for all $n$, since the procedure requires solving algebraic equations of order 3 and higher. The technical difficulty can be overcome by reverting to parametric representation of $f(X)$ as was done previously in other contexts in this paper, but the limitations on invertibility require more attention as already noticed by (e.g.) Plebanski\cite{Plebanski1970} and Bronnikov\cite{Bronnikov2000,Bronnikov2001}. 

Since the cases $(-1)^n \gamma <0$ produce unphysical electric solutions, there are two distinct kinds of behaviors of $X(n,\gamma,Z)$ for physical solutions: $(-1)^n$ and $\gamma$ both negative, or both positive. If  $(-1)^n$ and $\gamma$ are both negative, $X(n,\gamma,Z)$ is a monotonically increasing function and thus inverting for $Z(n,\gamma,X)$ is clearly possible. If $(-1)^n$ and $\gamma$ are both positive, then $X(n,\gamma,Z)$ is monotonic through $X=Z=0$ only up to its first (and single) maximum at a certain $Z$-value, say $Z_{max}$. So the inversion which produces $Z(n,\gamma,X)$, can be defined only up to this $Z_{max}$. Both physical subclasses are covered by substituting $\gamma=|\gamma|(-1)^n$ into the expressions for $X(n,\gamma,Z)$ and its derivative, $X_{_Z}$:
\be
X(n,|\gamma|,Z)=Z(1+2|\gamma| (-Z)^{n-1})^2 \; ,\;\;\;
X_{_Z}=\left(1+2|\gamma| (-Z)^{n-1}\right)\left(1+2(2n-1)|\gamma| (-Z)^{n-1}\right)
\label{PEDFOF-X(Z)}.
\ee
So, it is obvious that for odd $n$, $X(n,|\gamma|,Z)$ is monotonically increasing and the defining function  $f(n,|\gamma|,X)$ of the 2nd order formalism will be defined for all $X$. On the other hand, if $n$ is even, $X(n,|\gamma|,Z)$ has two extremal points for positive $Z$ values: a maximum  at $Z_{max}=1/(2(2n-1)|\gamma|)^{1/(n-1)}$ and a minimum at $Z_{min}=1/(2|\gamma|)^{1/(n-1)}$. Thus, for even $n$, the inverse Legendre transform and therefore the defining function $f(n,|\gamma|,X)$ are limited to $-\infty < X < X_{max}$ where
\be
X_{max} = \frac{4(n-1)^2}{(2n-1)^2}\,\frac{1}{(2(2n-1)|\gamma|)^{1/(n-1)}}
\label{PEDFOF-nEven-terminalPoint}.
\ee

Since the transformation $X(n,|\gamma|,Z)$ cannot be inverted analytically to give explicit expression for $Z(n,|\gamma|,X)$, we are limited to the parametric representation for the defining function $f_{_{ Z^n}}(n,|\gamma|,X)$:
\begin{eqnarray}
\left\{
\begin{array}{rl}
& f_{_{Z^n}}(n,|\gamma|,Z)= -Z+2\left(2-\frac{1}{n}\right)|\gamma|(-Z)^n  \\ 
\\
& X(n,|\gamma|,Z)=Z(1+2|\gamma| (-Z)^{n-1})^2 \, \, \, , \\
\end{array} \right.
   \label{fPEDFOF2ndOrder}
\end{eqnarray}
where for even $n$, the variables $X$ and $Z$ are bounded from above  by $X_{max}$ and $Z_{max}$ which are defined above.

Since an explicit expression for $f_{_{ Z^n}}(n,|\gamma|,X)$ is not available, we will give here its behavior near the origin and at large $X$. The behavior near the origin is trivially seen to be linear which corresponds to the primary request of Maxwellian behavior at weak fields. That is: $ f_{_{ Z^n}}(n,|\gamma|,X)  \simeq -X$. For strong fields we take first the limit $X\rightarrow -\infty$ where we find for either even or odd $n$:   
\be 
 f_{_{ Z^n}}(n,|\gamma|,X)  \simeq \left(2-\frac{1}{n}\right)\left(\frac{(-X)^n}{2|\gamma|} \right)^{1/(2n-1)}
  \label{AsymptoticfPEDFOF2ndOrder}
\ee
A similar expression is easily found for the $X\rightarrow \infty$ behavior  for odd $n$ only:
\be 
 f_{_{ Z^n}}(n,|\gamma|,X)  \simeq -\left(2-\frac{1}{n}\right)\left(\frac{X^n}{2|\gamma|} \right)^{1/(2n-1)}
  \label{AsymptoticfPEDFOF2ndOrderPlus}
\ee

\begin{figure}[t!]
		\centering
		\includegraphics[width=0.48\textwidth]{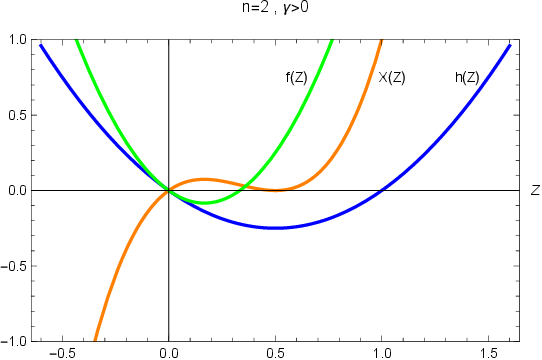} 
        \includegraphics[width=0.50\textwidth]{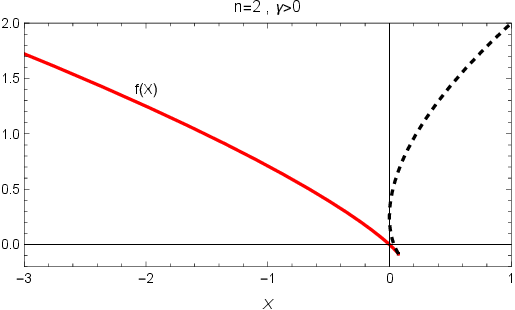}
		\caption{\small{The four functions that are involved in the Legendre transformation from the 1st order defining function $h(Z)$ of the  $Z^n$ model (with $n=2$) to its 2nd order counterpart, $f(X)$. The black dashed line at the right is the branch of $f(X)$ beyond the domain in which it is single valued.}}
		\label{PEDFOF-Legendre}
	\end{figure}

There is one additional point of interest which is the terminal point of $f_{_{ Z^n}}(n,|\gamma|,X)$ for even $n$ at $X=X_{max}$ given by Eq.\eqref{PEDFOF-nEven-terminalPoint}. Actually, this point turns out as a cusp point as can be found if we imagine to cross the point $Z=Z_{max}$ and move further. The reason is that the function $f_{_{ Z^n}}(n,|\gamma|,Z)$ has a local minimum at the same point $Z=Z_{max}$ where $X(n,|\gamma|,Z)$ has a local maximum. This is not specific to the $Z^n$ Lagrangian that is studied here, but a general feature of systems where $h(Z)$ is not monotonic such that $h_{_Z}$  vanishes for some finite $Z$. This can be seen easily from Eq.\eqref{hPEDFOFto2ndOrder}  which gives $X_{_Z}=h_{_Z} f_{_Z}$ for all $Z$. Therefore, the extremal points of $X(Z)$ and $f(Z)$ coincide, provided $h_{_Z}$ does not vanish too, which is true at least for the $Z^n$ Lagrangian. Actually, it is easy to calculate the slope of the $X$-dependent defining function $f_{_{ Z^n}}(n,|\gamma|,X)$ simply by $f_X = f_Z / X_Z$ at all points, and especially at the terminal point which turns out to be (for $n$ even):
\be
\frac{d}{dX} f_{_{ Z^n}}(n,|\gamma|,X_{max}) =-\frac{2n-1}{2(n-1)}
  \label{fPEDFOFSlope2ndOrder}
\ee
Fig.\ref{PEDFOF-Legendre}  demonstrates a graphic representation of the 4 functions that are involved in the (inverse) Legendre transform from the 1st order $ Z^n$ model to the 2nd order $ Z^n$ model: $ h_{_{ Z^n}}(Z)$,   $ X(Z)$,  $ f_{_{ Z^n}}(Z)$ and $ f_{_{ Z^n}}(X)$, for $n=2$. This represent the behavior for all even values of $n$. The situation for odd $n$ is simple and direct as we explained above and we skip the corresponding plots.

The main conclusion from this discussion about the equivalence between 1st and 2nd order formulations of NLED, is that the equivalence is not guaranteed to be 1 to 1 for all $X$ and $Z$. For $(-1)^n$ and $\gamma$ which are both positive, all the purely electric solutions that were found by the 1st order treatment go over  to the equivalent 2nd order formalism without any problem. However magnetic solutions (which we do not study in this work) are expected to encounter difficulties due to the partial equivalence in the $X>0$ (magnetic) domain. So, strictly speaking, insisting on full equivalence with second order NLED, this family of ``even'' PEDFOF models is ruled out as an acceptable NLED family. On the other hand, for the ``odd'' branch, $X(n,|\gamma|,Z)$ is monotonic and therefore invertible and is therefore viable. For example, all solutions - electric and magnetic are expected to be consistent with both formulations of the theory.  

\section{Palatini Nonlinear Electrodynamics}
\label{PNLED}
\setcounter{equation}{0}

So far, the 1st order NLED theories that we have considered were equivalent (at least partially) to 2nd order ones. However, we may give up this limitation and consider more general  1st order NLED Lagrangians which we name Palatini NLED (PNLED for short). These Lagrangians are determined by an arbitrary {\it defining function} of all 4 Lorentz scalars which may be constructed as bilinear products of $F_{\mu\nu}$ and $P_{\mu\nu}$: $Y$, $Z$ and $\Omega$ which were defined already, supplemented by $\Upsilon =F_{\mu\nu}\,^{*}P^{\mu\nu}$. We write therefore:
\begin{equation}
\mathcal{L}_{_{\rm PNLED}}=  \mathcal{K}(Z,\Omega , Y, \Upsilon)- J^\mu A_\mu\ ,
\label{LagNLED1stOrderGen}
\end{equation}
which gives rise to the following field equations:
\begin{equation}
-2\nabla_\mu \left(\mathcal{K}_Y P^{\mu\nu}+\mathcal{K}_\Upsilon \,^{*}P^{\mu\nu}\right)=J^\nu    \;\;\; , \;\;\;\;\
2 \left(\mathcal{K}_Z P^{\mu\nu}+\mathcal{K}_\Omega \,^{*}P^{\mu\nu}\right)+
\mathcal{K}_Y F^{\mu\nu}+\mathcal{K}_\Upsilon \,^{*}F^{\mu\nu}=0\ .
\label{FEqsNLED1stGen}
\end{equation}

It is easy to get convinced that the PNLED field equations cannot be mapped generically to the 2nd order ones. There are however some special cases (outside the minimal theory \eqref{LagNLED1stOrder3}) where it can be done. For example, if we choose $\mathcal{K}(Z,\Omega , Y, \Upsilon) = \kappa_1 (Z)/4-\kappa_2 (Y) /2$, we obtain the following field equations:
\begin{equation}
\nabla_\mu \left(\kappa'_{2}(Y) P^{\mu\nu}\right)=J^\nu    \;\;\; , \;\;\;\;\
 \kappa'_{1}(Z) P^{\mu\nu}-\kappa'_{2}(Y) F^{\mu\nu}=0\ .
\label{FEqsNLED1stGenExample}
\end{equation}
The homogeneous field equation determines the relation between $P^{\mu\nu}$ and $F^{\mu\nu}$ and can be solved in principle for $P^{\mu\nu}$ as a function of $F^{\mu\nu}$ (eliminating $Y$ even for nonlinear $\kappa_2 (Y)$) to be substituted in the source-full equation which will result a non-linear Maxwell equation of the form \eqref{NLED-FEqs} with $f_{\Xi}=0$. Still even this simple family of theories seems unstudied in the literature, probably because there is no Legendre transformation connecting this family of theories to 2nd order NLED.

 This is reminiscent of what occurs in GR, for which the standard second order formulation and the Palatini formulation are equivalent, while this ceases to be the case in higher order curvature theories. In the electromagnetic case the equivalence extends also to the non-linear regime, but it is limited to a certain family of theories as discussed in the previous sections.


\subsection{The PNLED-$Y^n$ model}\label{PNLEDYn}

It is well known that some NLED theories are able to mitigate the singularity of the electric field and its diverging total energy. The most popular is the Born-Infeld (BI) theory whose Lagrangian is given in Eq. \eqref{BI+HE}. The BI electric field $\mathcal{E}_{BI}$ in flat spacetime around a point charge turns out indeed to be finite everywhere, but it is still discontinuous at the location of the charge, say $r=0$. On the other hand it is possible to construct PNLED models where the electric field around a point charge is continuous everywhere, in particular $\mathcal{E}(0)=0$.

In order to demonstrate that, we study a simple PNLED Lagrangian where the nonlinearity originates from the $Y^n$ term of the defining function
$
 \mathcal{K}(Z,\Omega , Y, \Upsilon) = Z/4- Y/2 +\gamma Y^n/(2n)    \ .
$
More explicitly, we write the Lagrangian of the  ``$Y^n$ model'' as
\begin{equation}
\mathcal{L}_{Y^n}^{(1)}=\frac{1}{4}P^{\mu\nu}P_{\mu\nu}-\frac{1}{2}P^{\mu\nu}F_{\mu\nu}+\frac{\gamma}{2n}(P^{\mu\nu}F_{\mu\nu})^n  -J^\mu A_\mu \ ,
\label{LmodelYn}
\end{equation}
where $\gamma$ is a real parameter. It is clear that for $\gamma=0$, this reduces to the 1st order version of Maxwell's theory (compare (\ref{MaxwellFirst})) and $P_{\mu\nu}=F_{\mu\nu}$.

Now, variations with respect to $A_\mu$ and $P_{\mu\nu}$ give the modified source-full Maxwell equations and the modified $P$-$F$ relations:
\begin{equation}
\nabla_\mu\left[ \left(1-\gamma Y^{n-1}\right)P^{\mu\nu}\right]=J^\nu \; \;\; , \;\;\;\; P_{\mu\nu}=\left(1-\gamma Y^{n-1}\right)F_{\mu\nu} \ .
\label{FEqsModelYn}
\end{equation}
 In order to proceed, we need to rewrite Eqs.\eqref{FEqsModelYn} in terms of $F_{\mu\nu}$ only or $P_{\mu\nu}$ only. Contracting the second of~\eqref{FEqsModelYn} with $F^{\mu\nu}$ or $P^{\mu\nu}$ gives the following useful relations between $X$, $Z$ and $Y$:
 \be
 \gamma X Y^{n-1} + Y - X =0  \; \;\; , \;\;\;\;
 \gamma Y^{n} + Z - Y =0 \ ,
 \label{Eq_fdota_n}
 \ee
 Which result also the additional relations:
 \be
 W \equiv 1- \gamma Y^{n-1} = Z/Y = Y/X  \; \;\; , \;\;\;\;
 Y^2=XZ \ .
 \label{Eq_fdota_n-More}
 \ee
  So in principle the $P$-field and the $F$-field can be decoupled from each other. For a particular physical setup one can solve the first of~\eqref{FEqsModelYn} for $P_{\mu\nu}$ and then determine $F_{\mu\nu}$ from the second (assuring $dF=0$), or vice-versa. However, since algebraic equations of an arbitrary degree are involved, in practice solving the equations for $Y(X)$ or  $Y(Z)$ is expected to make the process cumbersome, unless a shortcut is found. Still the case $n=2$ is simple enough to allow a direct approach as is obvious from the trivial result of Eq.~\eqref{Eq_fdota_n}: $ Y=X/(1+\gamma X)$ which gives immediately the simple form of \eqref{FEqsModelYn}:
\begin{equation}
\nabla_\mu\left[ \frac{F^{\mu\nu}}{(1+\gamma X)^2}\right]=J^\nu \; \;\; , \;\;\;\;  P_{\mu\nu}=\frac{F_{\mu\nu}}{1+\gamma X} \;\;, \;\;\;\; n=2 \ ,
\label{FEqsModelYnSimp}
\end{equation}
which contains the modified source-full Maxwell equations written only in terms of $F_{\mu\nu}$. This simplification implies that this system is equivalent to a 2nd order NLED theory defined by Eq. \eqref{LagNLED2-general} with  $f(X)=X/(1+\gamma X)$. Flat space solutions of this $n=2$ system from a second order point of view were studied already by Kruglov \cite{Kruglov2014}, as well as the corresponding BH  metric \cite{Kruglov2015BH}.

The case $n=3$  is more complicated, but still manageable since Eq.~\eqref{Eq_fdota_n} become of second or third order in $Y$. But we will study it together with the general analysis of all $n\geq 3$.

But before that, we present the energy-momentum tensor of this kind of theories which we will need in flat space mainly for calculating the total energy and  in the presence of gravity for Einstein equations. Its general expression is given by:
\begin{equation}
T_{\mu\nu}=P_{\mu\alpha}P_{\nu}^{\ \alpha}-\left(P_{\mu\alpha}F_{\nu}^{\ \alpha}+P_{\nu\alpha}F_{\mu}^{\ \alpha}\right)\left(1-\gamma Y^{n-1}\right)-g_{\mu\nu}\mathcal{L}_{Y^n}^{(1)}
\label{EnergyDensity1stOrderNLED-FA_n}
\end{equation}
which simplifies (using also \eqref{Eq_fdota_n}-\eqref{Eq_fdota_n-More}) to
\begin{equation}
T_{\mu\nu}=-W^2 F_{\mu\alpha}F_{\nu}^{\ \alpha}-\frac{g_{\mu\nu}}{2}\left[ \frac{n-2}{2n}\,W -  \frac{n-1}{n} \right]Y
= - P_{\mu\alpha}P_{\nu}^{\ \alpha} -\frac{n-2}{4n}\,Z\,g_{\mu\nu} + \frac{n-1}{2n} Y g_{\mu\nu} \ .
\label{EnergyDensity1stOrderNLED-FA_n-Simp}
\end{equation}
For $n=2$ we get
\begin{equation}
T^{(n=2)}_{\mu\nu}= -\frac{F_{\mu\alpha}F_{\nu}^{\ \alpha}}{(1+\gamma X)^2}+\frac{g_{\mu\nu}}{4}\frac{X}{(1+\gamma X)}\ ,
\label{EnergyDensity1stOrderNLED-FA2}
\end{equation}
using~\eqref{FEqsModelYnSimp} to express it solely in terms of $F_{\mu\nu}$.

\subsection{Spherically-symmetric solutions of the $Y^n$ model in flat space}\label{Spherical-flat}
\textbf{\textit{n}=2}\\
We start with the simplest nonlinear case, $n=2$. In order to obtain the point charge fields of this model (in Minkowski spacetime) we use the same ansatz
$ A_t=A_t(r)$ as the only non-trivial component of the 4-potential $A_\mu$.  Then $F_{rt}(r)=\partial_r A_t(r)$ is the only non-trivial component of $F_{\mu\nu}$ and $X=-2F_{tr}(r)^2$. Therefore the inhomogeneous field equation of ~\eqref{FEqsModelYnSimp} gives (see also \cite{Kruglov2014})
\begin{equation}
 \frac{F_{tr}(r)}{\left[1-2\gamma F_{tr}(r)^2\right]^2}=\frac{Q}{r^2} \ ,
\label{modmax3}
\end{equation}
where $Q$ is an integration constant. This is an algebraic quartic equation for $F_{tr}(r)$. Although a closed form solution exists, it is long and unenlightening. On the other hand, it is much easier to inspect the LHS of \eqref{modmax3} in order to get an idea about the behavior of $F_{tr}(r)$. For that end we define $f_2 (F_{tr})=F_{tr}/(1-2\gamma F_{tr}^2)^2$. First of all we notice that for $\gamma>0$  $f_2 (F_{tr})$ diverges at $F_{tr}=1/\sqrt{2\gamma}$ which corresponds to $r=0$. This means that there are two branches of solutions: the first starts from the finite value of $F_{tr}(0)=1/\sqrt{2\gamma}$ at the origin and decreases monotonically to zero as $r\rightarrow \infty$ in a Coulomb-like way. As for the behavior near the origin, a little additional inspection shows that as
\begin{equation}
r\rightarrow 0 \quad \Rightarrow \quad F_{tr}(r)\simeq (2\gamma)^{-1/2}\left(1-\frac{r}{2(2\gamma)^{1/4}|Q|^{1/2}}\right)
\label{0lim}
\end{equation}
 The second branch
starts at the same central field value of $F_{tr}(0)=1/\sqrt{2\gamma}$, but is monotonically increasing with $r$ with asymptotic behavior of $F_{tr}(r)\simeq (r^2/Q \gamma^2)^{1/3}$. Notice also that in the branch of the increasing larger field, the two field functions, $F_{tr}(r)$ and $P_{tr}(r)$ have opposite signs which means opposite directions in space.

If we now turn to the second possibility of $\gamma <0$, we notice that $f_2 (F_{tr})$ does not diverge in this case, but it has a maximum at a certain value of $F_{tr}$. This maximum indicates a minimal possible value of $r$, say $r_{min}$ away from the origin, which means that the function $F_{tr}(r)$ is not defined in a ball of radius $r_{min}$ around the origin, but only for $r>r_{min}$. We will therefore consider this kind of solutions unphysical, or at least unfit for our purposes in this work.  The increasing field solutions with $\gamma >0$ will also be ignored in what follows, although they may have a physical role in  some circumstances.

One can now determine the corresponding  $P_{\mu\nu}$ from the $P$-$F$ relation of~\eqref{FEqsModelYnSimp} which we will do for the lower branch of the $\gamma>0$ solutions. The only non-trivial component is
\begin{equation}
P_{tr}=\frac{F_{tr}}{1-2\gamma F_{tr}(r)^2}\quad \stackrel{r\rightarrow 0}{\simeq}\quad (2\gamma)^{-1/4} \frac{|Q|^{1/2}}{r} \ ,
\label{modbiar}
\end{equation}
where~\eqref{0lim} was used. So $P_{tr}$ diverges at the origin, but more softly than  in Maxwell's theory, where it diverges as $1/r^2$. As a consequence, the total energy of the point particle in this model is actually \textit{finite} as we now discuss.

First we consider the energy density. Since only $F_{tr}$ and $P_{tr}$ are non-vanishing, we have for the energy density
\begin{equation}
\rho=T_0^0=\frac{F_{tr}^2(1/2+\gamma F_{tr}^2)}{(1-2\gamma F_{tr}^2)^2} \ .
\label{EnergyDensity1stOrderNLED-FASph}
\end{equation}
For $\gamma>0$ the energy density is positive.
 Moreover, a necessary condition for a finite total energy is a decreasing field, so we concentrate in the lower branch where $F_{tr}(r)$ decreases to zero asymptotically. For this branch we have
\begin{equation}
\rho=T_0^0\quad \stackrel{r\rightarrow 0}{\simeq}\quad  \frac{Q^2}{\sqrt{2\gamma}\,r^2} \ ,
\end{equation}
confirming that the total energy $E=4\pi\int r^2\rho dr$ is finite, despite the diverging energy density at the origin.

Actually, in this model the total energy can be represented by an analytical expression as a function of $\gamma$ and $Q$.
Since the explicit expression for  $F_{tr}(r)$ is cumbersome, the integration will be performed on $F_{tr}$ which for short we name $\xi$:
\begin{equation}
E=-4\pi\int_{0}^{\infty}  r^2 (\xi)\,  \rho(\xi)\,  \frac{dr}{d\xi} \, d\xi  = \frac{32 \cdot 2^{3/4} \pi }{15 }\; \frac{ Q^{3/2}}{\gamma^{1/4}}.
\label{TotalEnergy1stOrderNLEDSph_2}
\end{equation}

Since conventionally point particles couple to $A_\mu$ they ``feel'' the force as $F_{\mu\nu}$ and not as $P_{\mu\nu}$. This model is therefore advantageous with respect to the quartic GHE theory in 2nd order formalism, which has a finite field energy of a point charge, but a diverging force at the origin although weaker than the Coulomb one. We will now see that for $n\geq3$ the electric field at the origin will be continuous.\\ \vspace{-0.4cm}

\noindent \textbf{\textit{n} $\geq$ 3}\\
Analyzing the field around a point electric charge for all $n \geq 3$ cannot be done by repeating the $n=2$ procedure since solving analytically the algebraic relations \eqref{Eq_fdota_n}-\eqref{Eq_fdota_n-More} becomes impossible. Moreover, as we will see, for $n \geq 3$ the field strength $F_{tr}$ is not a monotonic function of $r$: rather it starts with $F_{tr}(0)=0$, achieves a maximum and decreases thereon. It is easy to see that it is the quantity $Y$ which may play the analogous role of  $F_{tr}$  as  an independent variable for our parametric solutions, and this will be done from now on. The inhomogeneous field equation of \eqref{FEqsModelYn} is readily integrated to $W(Y)P_{tr}=Q/r^2$ with $W = 1- \gamma Y^{n-1}$ from \eqref{Eq_fdota_n-More}. Assuming (without loss of generality), positive $Q$ and $P_{tr}$ we obtain the following parametric representation of the field $P_{tr}(r)$:
\begin{eqnarray}
\left\{
\begin{array}{rl}
P_{tr}= \sqrt{-W(Y)Y/2} \vspace{0.2cm}\\  
r = \left(\frac{\displaystyle{2Q^2}}{\displaystyle{-Y W^{3}(Y)}}\right)^{1/4}
\end{array} \right.
   \label{PradialYn}
\end{eqnarray}
 Similarly, the field strength $F_{tr}$ will be obtained using $F_{tr} = \sqrt{-Y/2W(Y)}$. Notice that for electric fields $Y\leq 0$. This imposes the correlation condition $\epsilon=(-1)^n sign(\gamma)=+1$ between $n$ and $\gamma$, which assures that the electric field of a point charge will be defined for all space and will decay asymptotically. We will see shortly that this same condition assures a positive definite energy density.

\begin{figure}[t!]
\begin{center}
{\includegraphics[width=0.49\textwidth]{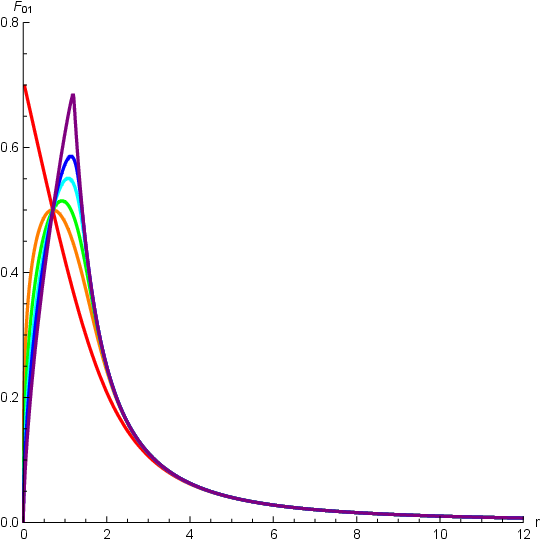}
\includegraphics[width=0.49\textwidth]{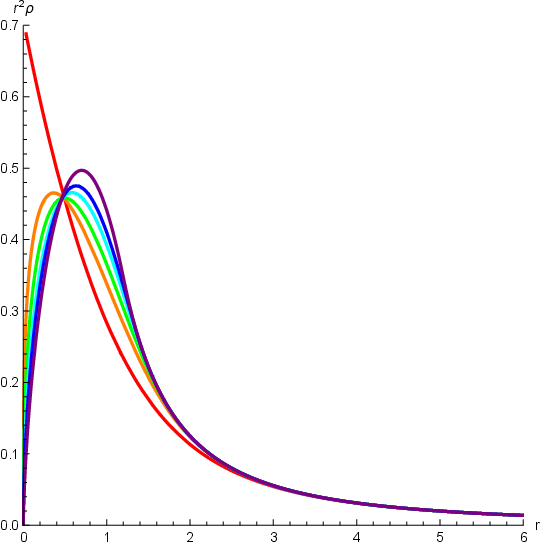}}
\caption{\small{Plots of the electric field $F_{tr} (r)$ (left) and $r^2 \rho(r)$  (right) for a $Q=1$ point charge for various values of $n$: $n=2,\;3,\;4,\;6,\;9,\;81$.  All curves start with zero at $r=0$ except for $n=2$ which decreases for all $r$. See discussion about the difference around Eq. \eqref{YnF01NearOrigin}.}}
\label{FigF01Profiles+E-DensityPNLEDYn}
\end{center}
\end{figure}

 The asymptotic behavior of $F_{tr}$ and $P_{tr}$ is easily found to be Coulombic. However, the behavior near the point charge (i.e. $r\rightarrow 0$) departs significantly from that:
 \begin{equation}
F_{tr}(r)\quad \stackrel{r\rightarrow 0}{\simeq}\quad
\frac{1}{(2^{n-1}  |\gamma|)^{\frac{2}{3 n-2}}}\left(\frac{r^2}{Q}\right)^{\frac{n-2}{3n-2}} \left(1-\frac{1 }{{(2^{4n-3}  |\gamma|)^{\frac{1}{3 n-2}}}}\left(\frac{r^2}{Q}\right)^{\frac{2(n-1)}{3 n-2}}\right)
\label{YnF01NearOrigin}
\end{equation}
Notice that $F_{tr}(r)$ does not diverge at the origin. For $n=2$ $F_{tr}(0)$ is finite while for all $n\geq 3$ it is zero.
 Fig. \ref{FigF01Profiles+E-DensityPNLEDYn}-left depicts some typical profiles of $F_{tr}(r)$, which shows a distinct peak near the origin for all $n > 2$, followed by a gradual decay to zero at large $r$. As $n$ increases, the peak of $F_{tr}(r)$ becomes more pronounced, indicating a stronger nonlinearity and further deviations from the Maxwellian field at large $n$. For $n = 2$, the electric field decreases monotonically with $r$, highlighting the unique behavior of this specific value of $n$ in this $Y^n$ model.

The corresponding $P_{tr}(r)$ behaves near the origin   differently than $F_{tr}(r)$, namely diverging at the origin like:
 \begin{equation}
 P_{tr}(r)\quad \stackrel{r\rightarrow 0}{\simeq}\quad \frac{1}{(2^{n-1}  |\gamma|)^{\frac{1}{3 n-2}}}\left(\frac{Q}{r^2}\right)^{\frac{n}{3n-2}} .
\label{FA3modelF01ProfileNearOrigin}
\end{equation}
As we will see now, this divergence is weak enough to render finite total energy for these solutions.
The energy-momentum components of spherically-symmetric electric field can be obtained from \eqref{EnergyDensity1stOrderNLED-FA_n-Simp} using  \eqref{Eq_fdota_n}-\eqref{Eq_fdota_n-More}. For the energy density we get
  \begin{equation}
\rho=T_0^0 = -\frac{3n-2}{4n}Z+\frac{n-1}{2n}Y =  -\frac{1}{4}Y + \frac{3n-2}{4n}\; \gamma Y^n ,
\label{T00YnModelAll_n}
   \end{equation}
which is positive definite as long as $\epsilon=(-1)^n sign(\gamma)=+1$. The pressure components are:
  \begin{equation}
T_r^r=T_0^0 \;\;\; , \;\;\;  T_\theta^\theta= T_\varphi^\varphi = \frac{1}{4}Y + \frac{n-2}{4n}\; \gamma Y^n .
\label{PressuresYnModelAll_n}
   \end{equation}
These are expressions in terms of $Y$, but together with $r(Y)$ of \eqref{PradialYn} their radial behavior may be obtained easily.
Fig.\ref{FigF01Profiles+E-DensityPNLEDYn}-right presents the energy density of these solutions, which are in accord with the behavior of the electric field.


The total field energy is obviously finite and is obtained by integration over $Y$, but before doing that, we pause to note that as in the previous sections, it is helpful to trade the self-interaction parameter $\gamma$ with an electric field parameter $\mathfrak{E}$ defined here such that $|\gamma| = 1/\mathfrak{E}^{2(n-1)}$ (no powers of 2 needed here). In addition, since we focus our attention on the physical branch of electric solutions where $Y<0$ we define a corresponding dimensionless variable which will absorb the minus sign as: $y=-Y/\mathfrak{E}^2=-|\gamma|^{1/(n-1)}Y$. Thus we replace the $Y$-integration by $y$-integration:
\begin{eqnarray}
E=4\pi\int_{0}^{\infty}  r^2\rho dr=-4\pi \mathfrak{E}^2\int_{0}^{\infty} dy\; \frac{dr}{dy}\; r^2 (y) \left[ \frac{1}{4}y + \frac{3n-2}{4n}\; y^n  \right]
\label{TotalEnergyYnModelAll_n}
\end{eqnarray}
which, using the dimensionless form of $r(y)$ from Eq.\eqref{PradialYn} is found to be: 
\begin{eqnarray}
E(\mathfrak{E},n,Q) = \frac{8\cdot 2^{3/4} \pi  Q^{3/2} \mathfrak{E}^{1/2} }{3} \cdot \frac{\Gamma\left(\frac{4n-3}{4 (n-1)}\right) \Gamma \left(\frac{5 n-6}{4(n-1)}\right)}{\Gamma \left(\frac{9}{4}\right) }
\label{TotalEnergyYnModelAll_n-Analytic} .
\end{eqnarray}

\subsection{Spherically-symmetric static black hole solutions of the $Y^n$ model}\label{Spherical-gravit}

We start our search for BH  solutions of this new PNLED theory with the self-gravitating version of the Lagrangian of Eq. \eqref{LmodelYn}:
\begin{equation}
\mathcal{L}=\frac{1}{2\kappa}R+\frac{1}{4}P^{\mu\nu}P_{\mu\nu}-\frac{1}{2}P^{\mu\nu}F_{\mu\nu}+\frac{\gamma}{2n}(P^{\mu\nu}F_{\mu\nu})^n \; .
\label{EqSelf-Grav-NLED-FA_n}
\end{equation}
A simple inspection reveals that for spherically-symmetric electrostatic solutions we have the same situation as with the two previous models: the spherically symmetric metric of Eq. \eqref{EqStaticSphericalMetric} is still valid, as well as the same flat-space $Y^n$ field tensor $P_{\mu\nu}$ of \eqref{PradialYn} and the associated $F_{\mu\nu}$. The corresponding components of $T_\mu^\nu$ of  Eq.\eqref{EnergyDensity1stOrderNLED-FA_n} or \eqref{EnergyDensity1stOrderNLED-FA_n-Simp} reduce in this case to \eqref{T00YnModelAll_n}-\eqref{PressuresYnModelAll_n} resulting a single metric function by $u(r)=f(r)$ and two independent $T_\mu^\nu$ components.   
 Therefore, we are left with only one field equation for the single metric function $f(r)$. The second Einstein equation will be used as a check on the solutions. We will not treat here the case $n=2$ separately (see\cite{ Kruglov2014,Kruglov2015BH} for partial results), but write the field equations for all $n\geq2$  still imposing   $\epsilon=(-1)^n sign(\gamma)=+1$:
\begin{eqnarray} 
\frac{1}{r^2 }\frac{d}{dr}r(1-f)=\kappa \left(-\frac{1}{4}Y + \frac{3n-2}{4n}\; \gamma Y^n \right) 
\quad \Rightarrow \quad
\frac{dm}{d\varrho} = \frac{\varrho^2 }{2} \left(\frac{1}{4}y + \frac{3n-2}{4n}\; y^n \right)
\label{FEqs-Grav-NLED-FA_n}
\end{eqnarray}
where a dimensionless radial variable and mass function were defined by rescaling with the length parameter $\ell = 1/\sqrt{\kappa \mathfrak{E}^{2}}$.
Change of variables from $\varrho$ to $y$ gives the following equation for $m(y)$:
\begin{equation}
\frac{dm}{dy} = \frac{\varrho^{2} (y)}{2}  \left(\frac{1}{4}y + \frac{3n-2}{4n}\; y^n \right)\frac{d\varrho}{dy} 
\label{EinsteinEqMass-NLED-FA_n}
\end{equation}
where $\varrho(y)$ is given by the dimensionless form of the flat space expression from \eqref{PradialYn}:
\begin{equation}
\varrho = \left(\frac{2q^2}{y(1+y^{n-1})^3}\right)^{1/4}
\label{Rho-y},
\end{equation}
with $q=\kappa\mathfrak{E}Q$.  Integration gives the following (dimensionless) mass function:
\begin{eqnarray}
m(y)=\frac{q^{3/2}}{ 2^{1/4}\cdot 15(n-1)}\left[ \frac{n (17 n-49)+(10+n (27 n-101)) y^{n-1}-32 n y^{2 (n-1)}}{4 n \left(1+y^{n-1}\right)^{9/4}}y^{1/4} +
\right. \nonumber \\ \left.
\frac{8} {y^{(n-2)/4}} F\left(\frac{1}{4},\frac{n-2}{4 (n-1)},\frac{5 n-6}{4 (n-1)},-\frac{1}{y^{n - 1}}\right)
\right]+m_0  =  m_{field} (y) +m_0
   \label{DmlssMassFunction-NLED-FA_n},
\end{eqnarray}
where $m_0$ is an integration constant which plays a role of a point mass at the origin, since $\lim_{y\rightarrow \infty} m(y)=m_0$. We introduced here for further use the function $m_{field} (y)$ which represents the accumulated field energy from the origin up to the sphere of a certain value of $y$. The total BH mass is obtained by taking the  limit at the other end:  
\begin{equation}
m_{_{BH}} = \lim_{y\rightarrow 0} m(y) = m_0 + \frac{2^{3/4} q^{3/2} }{3} \cdot \frac{\Gamma\left(\frac{4n-3}{4 (n-1)}\right) \Gamma \left(\frac{5 n-6}{4(n-1)}\right)}{\Gamma \left(\frac{9}{4}\right) } = m_0+\overline{m}_{field}
\label{BHmass-NLED-Yn},
\end{equation}
This is of course just the sum of $m_0$ and the field energy obtained already in  \eqref{TotalEnergyYnModelAll_n-Analytic}. 

The metric function $f(\varrho)$ will be given in parametric form by the condensed expression: 
\begin{equation}
\left\{
\begin{array}{rl}
&f(y)= 1-\frac{\displaystyle{2 m (y)} }{\displaystyle{\varrho(y)}} =  1-\frac{\displaystyle{2 (m_{field} (y) + m_{_{BH}}- \overline{m}_{field})}}{\displaystyle{\varrho(y)}} 
\vspace{0.2cm}\\ 
&\varrho(y) = \left(\frac{\displaystyle{2q^2}}{\displaystyle{y(1+y^{n-1})^3}}\right)^{1/4}  . \\
\end{array} \right.
\label{fSolBHDmlss-NLED-FA_n}
\end{equation}

\begin{figure}[t!]
\begin{center}
\includegraphics[width=0.49\textwidth]{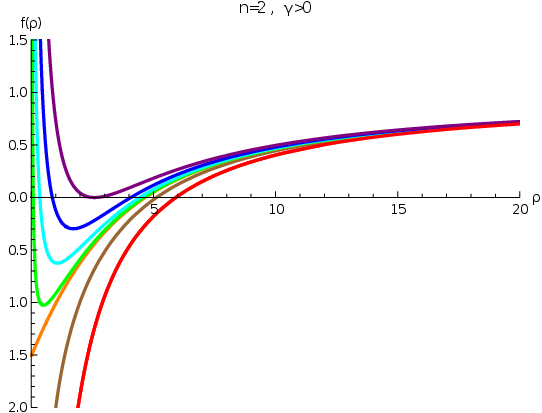}
\includegraphics[width=0.49\textwidth]{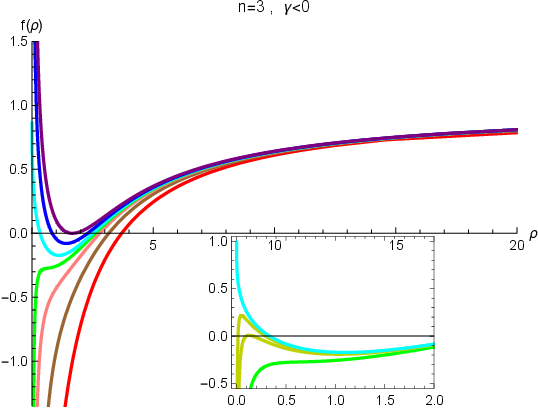}
\caption{\small{Plots of the metric function $f(\rho)$  of $Y^n$ BHs.
 Left: $n=2$, $\gamma>0$ , $m=3$, $q=1,\;3,\;3.550205,\;3.6,\;3.75,\;4.0,\;4.40353$;
 Right:  $n=3$, $\gamma<0$, $m=2$ ; $q=1.5,\;2.3,\;2.6,\;2.7,\;2.737,\;2.745,\; 2.74893,\;2.82,\;2.899366$.
The charge increases upwards. Note especially: (i) the intermediate solutions with a finite $f(0)$ in both cases and their different behaviour; (ii) the 3-horizon solutions for $n=3$ - see insert.}}
\label{Fig-f-NLED-FA_nNEW}
\end{center}
\end{figure}

Fig. \ref{Fig-f-NLED-FA_nNEW} presents typical profiles of the metric function $f(r)$ for several mass values for a fixed charge for $n=2$ (left) and $n=3$ (right). The features of the $n=3$ BHs are also typical to all higher $n$-values. The $n=2$ BHs are rather exceptional as is already evident from their electric field depicted in Fig.\ref{FigF01Profiles+E-DensityPNLEDYn}.

It is obvious that for all $n\geq 2$ there are two distinct types of solutions characterized by the relation between charge and mass, or ore concretely by the ratio $\overline{m}_{field}/m_{_{BH}}$. For a given $M$, small charge BHs have a S-like behavior with a single horizon. For higher charges, above a certain critical charge (for which $\overline{m}_{field}=m_{_{BH}}$), a second horizon appears and the behavior becomes RN-like. The RN-like behavior  extends up to a maximal value of charge, above which the solutions exhibit naked singularities.

Between these two families, there exists  (unlike the case of the analogous solutions in the 2nd order formalism), intermediate solutions which have  for all $n\geq 2$ a non-divergent $f(0)$. Although $F_{tr}$ is also finite (or even zero) at the origin for the $n\geq 2$ solutions, the energy density diverges there, so these solutions are not regular BHs. Since they correspond to  the conditions $\overline{m}_{field}=m_{_{BH}}$, or  $m_0=0$, they may be thought as solitonic BHs which are ``made of'' a non-trivial finite energy \textit{regular} field configuration of the self-interacting vector field. Notice however the difference between the near origin behavior of the intermediate  $n=2$ and the $n\geq 3$ solutions: First of all, the $n=2$ ones are S-like while the others are RN-like. In addition, the  $n=2$ solution approaches a negative charge-dependent $f(0)$ with a positive finite slope. On the other hand, all $n\geq 3$ intermediate BH solutions tend to $f(0)=1$ with a diverging slope. All this is a realization of the following result for the small $\rho$ behavior of $f(\rho)$ with $m_0 =0$:
\begin{equation}
f(\rho) \simeq  1-\frac{(3 n-2)}{2^{\frac{5 n-4}{3 n-2}}n}\left[\frac{(3 n-2) q^{\frac{2 n}{3 n-2}} }{5 n-6}\rho ^{\frac{2 (n-2)}{3 n-2}}
-\frac{5 (7 n-6) q^{\frac{2}{3 n-2}} }{2^{\frac{7n-5}{3 n-2}}(9 n-10)} \rho ^{\frac{2 (3 n-4)}{3
   n-2}}\right]
\label{YnBHnearOrigin}.
\end{equation}

For $n\geq 3$ a new kind of solutions emerges in the vicinity of the intermediate solutions  (slightly smaller charge) whose main feature is a third horizon such that $f(\rho)\rightarrow -\infty$. Still it may not be classified as S-like, but deserve a third class of its own.
So, by increasing the charge of a $n\geq 3$ BH, the single horizon S-like solutions turn smoothly to 3-horizon BHs and only after further charge  increase the 2-horizon RN-like solutions appear. 

 Since the solutions of this $Y^n$ model have different characteristics for $n=2$ and $n\geq 3$, we will keep presenting all the plots throughout this section for both of them. Usually we will use  $n=3$ to represent the whole $n\geq 3$ family. We will keep imposing always $\epsilon=+1$. 

\begin{figure}[b!]
\begin{center}
\includegraphics[width=0.49\textwidth]{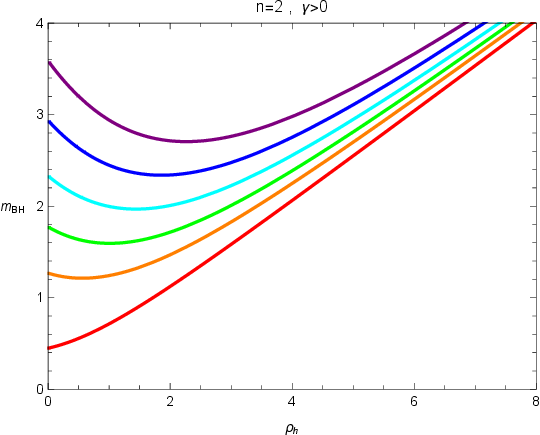}
\includegraphics[width=0.49\textwidth]{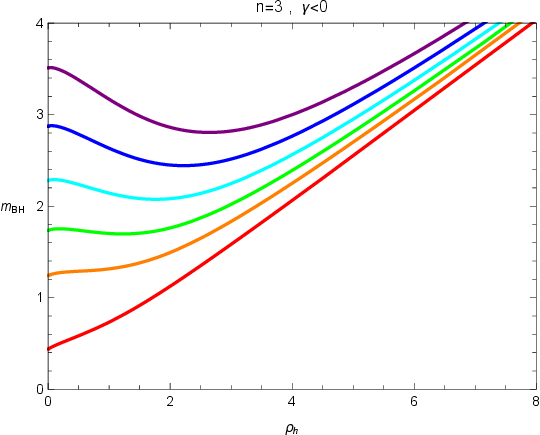}
\caption{\small{Plots of the BH mass vs the horizon radius $\varrho_h$ for  $n=2$ (left) and $n=3$ (right). The mass curves correspond several values of the dimensionless charge parameter: $q = 1,\, 2,\, 2.5,\, 3,\, 3.5,\, 4$. }}
\label{Fig-mBH-rhoh}
\end{center}
\end{figure}

Finally for this subsection, we will get the relation between the BH mass, charge and horizon parameter, also in parametric form, by solving for $m_{_{BH}}$ the equation $f(y_{_h})=0$, using \eqref{DmlssMassFunction-NLED-FA_n}, \eqref{BHmass-NLED-Yn} and \eqref{fSolBHDmlss-NLED-FA_n}. Since the relevant expressions are quite cumbersome, we will not give them in full-length, but write the expression for BH mass in the following form (cf \eqref{NLED2MassvsHorizons-compact}):
\begin{equation}
m_{_{BH}}=\overline{m}_{field}+\frac{1}{2}\varrho(y_{_h})-m_{field}(y_{_h})
\label{BHmassvsQ+yhPNLEDYnDmlss}
\end{equation}
where $\overline{m}_{field}$ is the total field energy given in the second term of  Eq.\eqref{BHmass-NLED-Yn}, $\varrho(y_{_h})$ is the horizon radius and $m_{field}(y_{_h})$ is the horizon value of the field mass function of Eq.\eqref{DmlssMassFunction-NLED-FA_n}. Fig. \ref{Fig-mBH-rhoh} depicts the dependence of the BH mass on the horizon radius with several values of charge, for the two lowest values of $n$.The $n>3$ curves are generally similar to the $n=3$ ones. These plots can be used easily (more easily than the fixed mass plots of Fig.\ref{Fig-f-NLED-FA_nNEW}) to infer the  characteristics of fixed charge  BHs. So they summarize more fully the main features of the BH type studied here. Like the GHE BHs of Sec. \ref{GHE}, 
 a BH with a given large enough charge, may be a 2-horizon RN-like BH, but if its mass increases (say, by accreting neutral particles), it turns into a single-horizon S-like BH. This is described by the upper curves which exhibit a minimal BH mass for both $n=2$ and $n=3$ panels of Fig. \ref{Fig-mBH-rhoh}. However, unlike the GHE diagram of Fig. \ref{Fig-BHmass+ChargeVsrH-NLED2nd} (left), in this case there exist for low enough  $q$ curves with monotonically increasing BH mass which represent a family of purely S-like solutions. The fact that these S-like solutions do not contain extremal BHs, has a crucial effect on their thermodynamic aspect as we will see in the next subsection. In addition to the above, one may notice in the $n=3$ panel  of Fig. \ref{Fig-mBH-rhoh} (but seen for all $n\geq 3$) a small region of intermediate charge values containing curves with both a local maximum and minimum, represents a BH family which contain 3-horizon BHs too. These BHs have also some peculiar thermodynamic features. 

\subsection{Thermodynamical properties}
The Hawking temperature for this kind of BHs can be calculated as in Sec.\ref{GHEBHs-Thermodynamics} using the $00$ Einstein equation \eqref{FEqs-Grav-NLED-FA_n} for an easy explicit expression for $f'(r_h)$:
\begin{eqnarray}
 \frac{\ell k_B}{\hbar} T_H=\tau_H= \frac{1}{4 \pi}\left[\frac{1}{\varrho_h}-\varrho_h \left(\frac{1}{4}y_{_h} + \frac{3n-2}{4n}\;  y_{_h}^n \right) \right]
\label{BHTempNLED1stOrderYn}.
\end{eqnarray}
\begin{figure}[b!]
		\centering
		\includegraphics[width=0.49\textwidth]{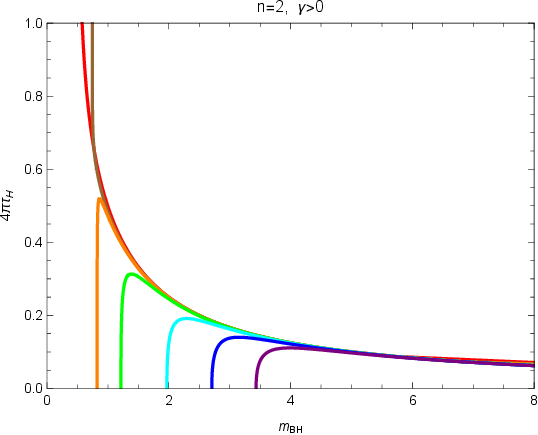} 
        \includegraphics[width=0.49\textwidth]{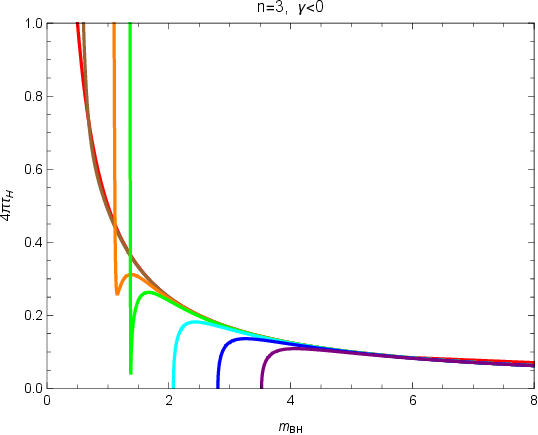}
		\caption{\small The Hawking temperatures of the  $Y^n$ mo BHs as a function of the mass for several values of charge. Left: $n=2$ and   $q = 1, 1.4, 1.5, 2, 3, 4, 5$;  Right: $n=3$ and $q = 0.25, 1, 1.8, 2.1,  3, 4, 5$. Notice the deep temperature minima for $n=3$ which correspond to S-like BH solutions with charges just below the $q$-value of the intermediate solution which separates the RN-like family from the S-like one.}
		\label{fig:Temp-mBH-Sec6}
\end{figure}

\begin{figure}[t!]
\centering
\includegraphics[width=0.48\textwidth]{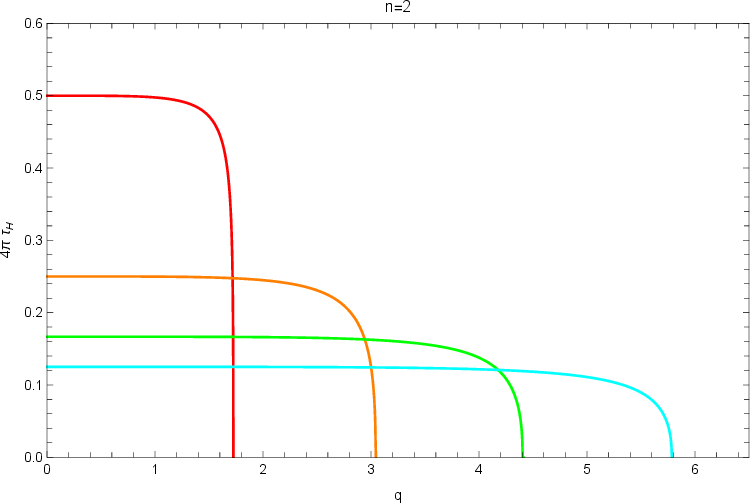} 
\includegraphics[width=0.48\textwidth]{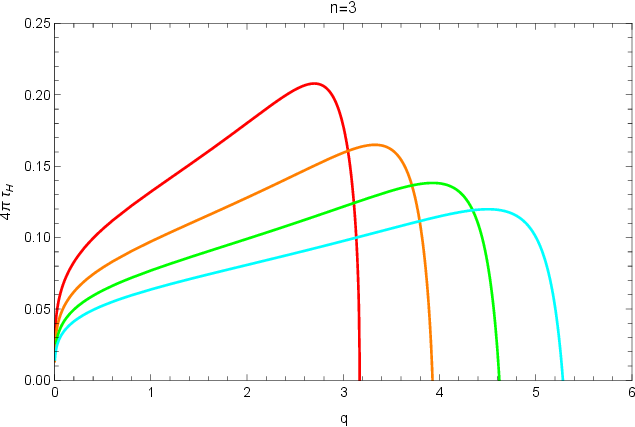} 
\caption{\small The Hawking temperatures of the  $Y^n$ model BHs as a function of the charge for several values of mass. Left: $n=2$ and   $m = 1,2,3,4$;  Right: $n=3$ and $m = 3,4,5,6$.}
\label{fig:TempQ-Sec6}
\end{figure}

This should be supplemented by the expression for $\varrho_h=\varrho(y_h )$ from Eq.\eqref{Rho-y} to get a parametric description of $T_H (Q, r_h)$.

The dependence on $r_h$ (or $\varrho_h$) of the temperature of a BH with a given charge can be obtained in a parametric form, using the above equation (\ref{Rho-y}). We leave out this plot and show only the mass dependence of the temperature in Fig. \ref{fig:Temp-mBH-Sec6}  and its charge dependence in Fig. \ref{fig:TempQ-Sec6}. In the left panel of Fig. \ref{fig:Temp-mBH-Sec6} for $(n = 2, \gamma > 0)$, the temperature curves exhibit a typical behavior for charged black holes for large enough charge values. For each fixed charge, the temperature increases rapidly from  $T=0$ at the EBH point, reaches a peak, and then decreases monotonically as the mass increases, approaching zero in the limit of large mass. Higher charge values  shift the temperature peak to larger masses but lower temperature, indicating the stabilizing effect of charge. All this is quite common, but the low charge behavior is significantly different due to the existence of the S-like solutions. The two upper curves of the left panel correspond to S-like BHs which have no maximal temperature, but keep increasing as the mass decreases. The right panel exhibit similar structure, but presents a new feature for $n\geq 3$ which is the appearance of minima at the temperature curves of the S-like BHs. These minima result from the fact that just before the $f(\varrho_h)$ curves acquire the 3-horizon behavior, the  \textit{second} derivative  $\partial^2 f(\varrho_h) \partial \varrho_h^2$ becomes negative at a certain interval causing a change in the slope of the temperature curves. This will also be reflected in additional structure in the heat capacity of these families of solutions.

\begin{figure}[t!]
		\centering
\includegraphics[width=0.49\textwidth]{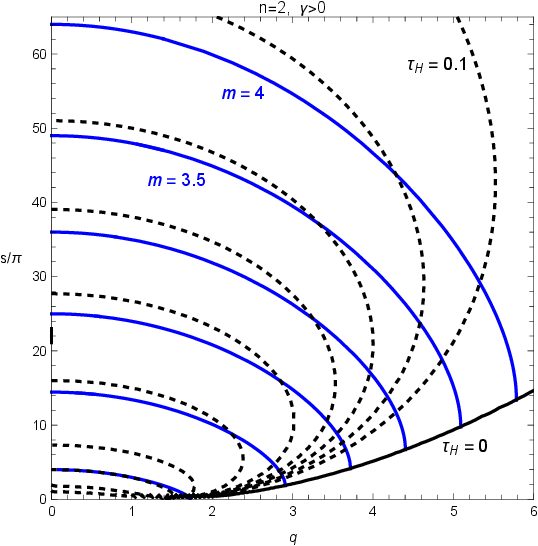}
\includegraphics[width=0.49\textwidth]{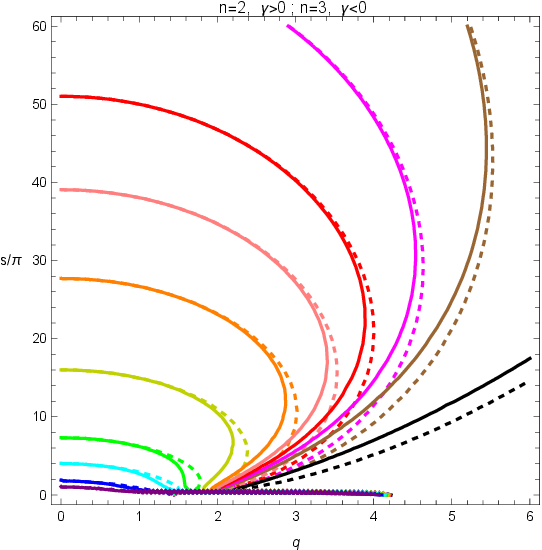} 
\caption{ \small Curves of the $Y^n$-BH  entropy as a function of charge for various values of mass  and  temperature in the cases $n=2$ and $n=3$. Left: Lines of constant  $\tau_H$ and constant $m_{BH}$ in the $q$-$s$ plane, for $n=2$. Mass values: $m_{_{BH}}=1.0, 1.9, 2.5, 3.0, 3.5, 4.0$. Temperatures: $4\pi\tau_H = 0, 0.10, 0.12, 0.14, 0.16, 0.19, 0.25,$ $ 0.37, 0.50, 0.75, 1.00$.  Several numerical values are added in order to indicate the way $m_{_{BH}}$ and $\tau_H$ increase.
Right: Isotherms for both $n$-values. The $\tau_H$-values are the same as in the left panel for both $n$'s. }       
\label{fig:entropy-Sec6}
\end{figure}

The Hawking-Bekenstein entropy of the BHs of this model is written in the dimensionless version as the $y_{_h}$-dependent function  
\begin{equation}
s= \pi \left(\frac{\displaystyle{2q^2}}{\displaystyle{y_{_h}(1+y_{_h}^{n-1})^3}}\right)^{1/2}
\label{BH-Entropy-Yn},
\end{equation}
and its main properties are shown in Fig. \ref{fig:entropy-Sec6} using $\varrho(y)$ from Eq.\eqref{fSolBHDmlss-NLED-FA_n} for a parametric representation together with change of variables for plotting in the $q$-$s$ plane the lines of constant mass and of constant temperature. We chose to present the constant temperature and constant mass lines  for $n=2$ only  (left panel - cf Fig. \ref{fig:entropy}). In the right panel we show for both $n=2$ and $n=3$, only the constant temperature lines in the $q$-$s$ plane  for the sake of a direct impression of the general similarity for the large entropy region, as well as the differences for lower $s$ values.  

The heat capacity is also calculated in a dimensionless form by a parametric representation, namely: 
\begin{equation}
 c = \frac{\partial m }{\partial \tau_{H}} = \frac{\partial m / \partial y_h }{\partial \tau_{H} / \partial y_h}
\label{BH-HeatCapacityYn}\, .
\end{equation}
Using the compact equation \eqref{BHmassvsQ+yhPNLEDYnDmlss} for the BH mass and the field equation \eqref{EinsteinEqMass-NLED-FA_n} one finds the following expression for the rescaled heat capacity of the $Y^n$ black holes
\begin{eqnarray}
\frac{c}{4\pi}=-\frac{q \left(2 \sqrt{2}\, n \left(1+y_h^{n-1}\right)^{3/2}-q y_h^{1/2} \left(n+(3 n-2)
   y_h^{n-1}\right)\right)}
{\sqrt{2}\, y_h^{1/2} \left(y_h^{n-1}+1\right)^{3/2} \left(2 \sqrt{2}\, n
   \left(1+y_h^{n-1}\right)^{3/2}-q y_h^{1/2} \left(3
   n+(n+2) y_h^{n-1}\right)\right)}
   \label{Yn-HeatCap}.
\end{eqnarray}

\begin{figure}[t!]
		\centering
		\includegraphics[width=0.48\textwidth]{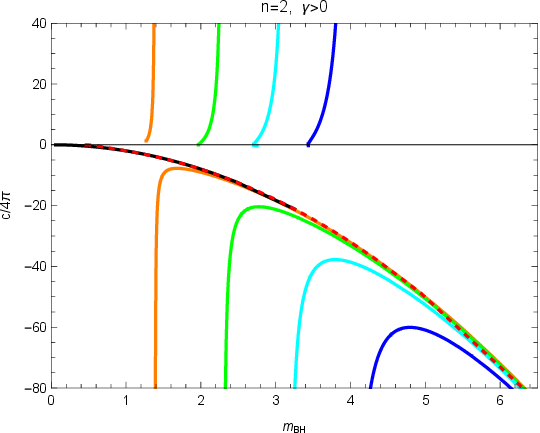} 
        \includegraphics[width=0.48\textwidth]{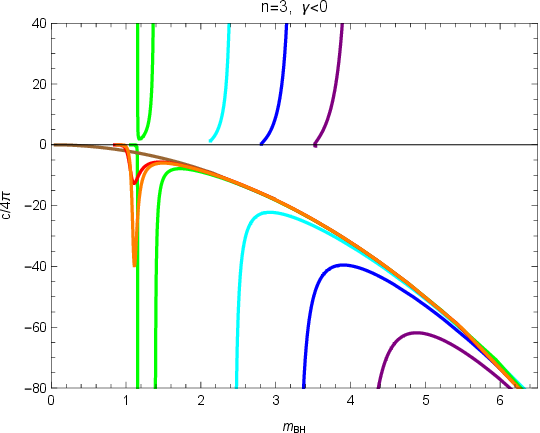}
		\caption{\small  The heat capacity as a function of mass of BHs of the $Y^n$ model with $n=2$ and $n=3$. Left: $n=2, \gamma>0$. Heat capacity  for $q = 0.25, 1, 2, 3, 4, 5$; Right: $n=3, \gamma<0$. Heat capacity for  $q = 0.25, 1.55, 1.59, 1.8, 3, 4, 5.$. Notice the Schwarzschild-like curves for low charges in both $n$-values. Notice also the new structure for $n=3$. }
		\label{fig:HeatCapacitymBH-Sec6}
\end{figure}

Figure \ref{fig:HeatCapacitymBH-Sec6} present the main features of the heat capacity of this kind of BHs as the dependence of $c$ on the mass and charge. The pattern contains some conspicuous differences with respect to  the 2 previous types discussed in the previous sections. The first is the fact that the Heat capacity curves split into 2 types: (i) high charge RN-like with the singularities between the positive and negative regions, (ii) S-like curves at small charges which are always negative and decrease with $m$ indefinitely.  
Moreover, unlike the previous model, now there are major differences between the $n=2$ and $n\geq3$ classes. For instance,  discontinuities of the heat capacity appear as expected already due to the existence of points of local maxima and minima of the temperature. There appear also new branches of continuous and negative heat capacity, corresponding to the domain of parameter space where the temperature has no local maxima.

\section{Conclusions and Summary}
\label{conclusion}
In this paper we have demonstrated that NLED theories may be generalized using 1st order formulation beyond the ``conservative'' existing type of theories, if the condition of equivalence to 2nd order formulation is relaxed. The new 1st order formalism presented here, named ``Palatini Nonlinear Electrodynamics'' may be used also to discover new ``conservative'' NLED models which are equivalent to 2nd order NLED ones, but do not have a simple explicit 2nd order representation, while they can be represented simply within 1st order.

A specific family of models of the PNLED was constructed in Sec.\ref{PNLED} and its spherically-symmetric electrostatic solutions were analyzed intensively. This family of ``$Y^n$ models" turns out to exhibit a rich family of charged solutions, both self-gravitating or in flat spacetime.  One of the most intriguing aspect of this new family is the fact that the same model allows for several kinds of BH solutions that may transform to each other in processes which e.g. conserve electric charge but allow mass changes. These transformations  may change the causal structure of the BHs by changing the number of horizons - a process that deserves further study in order to gain deeper understanding of the nature of these BHs.

An additional interesting result already in flat spacetime, is the spherically-symmetric electric field of the ``$Y^n$ models'' with $n\geq 3$ which is not only finite  at the origin as the BI field (which is though discontinuous), but vanishes at the origin, thus rendering the electric field continuous there. The existence of the point charge is reflected by the discontinuity of the first derivative. The total field energy is of course finite.

As a ``background'' for better appreciation of the $Y^n$ spherically-symmetric solutions, we presented two other families of NLED solutions: the first is the electric point charge of the  generalized Heisenberq-Euler theory of Equation \eqref{EqSelf-Grav-NLED-2ndOrder} treated in 2nd order formalism. Its electric field diverges at the origin, but its total field energy is finite. The corresponding BH solutions are either single-horizon S-like, or 2-horizon RN-like BHs. The second family is polynomial electrodynamics with a similar action (Eq.\eqref{EqSelf-Grav-NLED-Z_n})  treated in 1st order. Although similar and often considered equivalent to the GHE theory for weak fields, in general there is no equivalence whatsoever, but considerable differences: The electric field of a point charge diverges at the origin in such a way that the total field energy also diverges. The corresponding BH solutions are all RN-like with similar characteristics. 

It is worth noting that all 3 models studied in this paper may be solved analytically. One directly in terms of $r$-dependent functions, and the 2 others using the mathematical approach of parametric representation of the solutions, thus allowing relatively simple expressions for (and among) the main physical parameters of the models, like mass, charge, horizon radius, temperature, entropy etc. 

The solutions presented here were limited to be purely electric, but it is already clear from preliminary work done by us, that corresponding magnetic solutions exist as well, with new interesting features. Dyonic solutions are an immediate extension, which will ``thaw'' the ``$F$-$F$-dual'' terms in the Lagrangians, thus expected to produce new kinds of compact objects. 

A related interesting issue is the existence of regular black holes in the new PNLED theory which was shown here to contain in its simple $Y^n$ model BH solutions with a globally regular electric fields. This could imply that a slight modification of the $Y^n$ model will produce new regular BHs which are always welcome.

Further research directions emanating from the present project are the analysis of geodesics of the new PNLED BH geometries - both null and timelike. The former correspond to  the optical features of these BH solutions like light rings, shadows and gravitational lensing and it is especially well-timed at the present era of precision BH observations which have produced already the ground-breaking results of EHT \cite{EHT_M87,EHT_SgrA} as well as other telescopes and detectors. Actually, work in this direction is currently in progress by the authors of this paper. The latter ``complementary'' path of timelike (spatially closed) geodesics of these BH geometries is an additional way of revealing  the idiosyncratic features of  PNLED BHs from the behavior of orbiting ``satellites'' and other massive particles. There are of course two very different BH mass scales that should be considered: the BH mass scales of 10-100 solar masses in the LIGO-VIRGO-KAGRA \cite{KAGRA:2021duu} ``window'' and the supermassive BHs studied by the EHT. 

\vspace{0.4cm}
\noindent{\textbf{Acknowledgments}} 

Useful and illuminating discussions with Carlos Herdeiro, Jutta Kunz, Wilhelm Mensing, Syksy Rasanen and Damien Eason are gratefully acknowledged. B.P. and Y.V. would like to thank for financial support by the Israeli fund for scientific regional cooperation and the Research Authority of the Open University of Israel. The work of B. P. is partially supported by Sabanc{\i} University, Faculty of Engineering and Natural Sciences. A. {\"O}. and B. P. would like to acknowledge the contribution of the COST Action CA21106 - COSMIC WISPers in the Dark Universe: Theory, astrophysics and experiments (CosmicWISPers) and the COST Action CA22113 - Fundamental challenges in theoretical physics (THEORY-CHALLENGES). H. H. is supported by the National Natural Science Foundation of China (NSFC) Grant No. 12205123, and Jiangxi Provincial Natural Science Foundation, Grant No. 20232BAB211029.

\end{document}